\documentclass[11pt]{article}

\usepackage{graphicx} 
\usepackage{amsmath}
\usepackage{amsfonts}
\usepackage{amsthm}
\usepackage{bbm}
\usepackage{multirow}
\usepackage{leftindex}
\usepackage{setspace}
\usepackage{array}
\usepackage[colorlinks=true, linkcolor=green, citecolor=blue, urlcolor=blue]{hyperref}
\usepackage{yfonts}
\usepackage{booktabs}
\usepackage{cleveref}
\usepackage{comment}
\usepackage{todonotes}
\usepackage{amssymb}
\usepackage{color}
\usepackage{subcaption}
\usepackage{float}
\usepackage[round]{natbib}
\usepackage[hmargin=2cm,vmargin=2cm]{geometry}

\newcommand{\feeLTsells}{\mathfrak{p}}
\newcommand{\feeLTbuys}{\mathfrak{m}}
\newcommand{\exratesell}{Z_{+}}
\newcommand{\exratebuy}{Z_{-}}

\newcommand{\depth}{p}

\newcommand{\oraclepricestochastic}{S_{t}}

\newcommand{\fees}{\mathfrak{f}}

\newcommand{\Cash}{\mathfrak{C}}
\newcommand{\cash}{\mathfrak{c}}

\theoremstyle{plain}
\newtheorem{theorem}{Theorem}[section]

\theoremstyle{definition}

\newtheorem{remark}[theorem]{Remark}

\theoremstyle{remark}
\numberwithin{equation}{section}
\newtheorem*{empty*}{}

\title{
Competition between DEXs through Dynamic Fees
}

\author{Leonardo Baggiani\thanks{Corresponding author: leonardo.baggiani@warwick.ac.uk}\,\,\thanks{Department of Statistics, University of Warwick}
\and
Martin Herdegen\thanks{Department of Mathematics, University of Stuttgart}
	\and 
Leandro S\'anchez-Betancourt\thanks{Mathematical Institute and Oxford-Man Institute of Quantitative Finance, University of Oxford}
	}

\newcommand{\R}{\mathbb{R}}

\newcommand{\ind}{\mathbbm 1}

\newcommand{\dd}{\,\mathrm{d}}

\renewcommand{\epsilon}{\varepsilon}

\begin{document}

\maketitle

\begin{abstract}
We find an approximate Nash equilibrium to a game between decentralised exchanges (DEXs) that compete for order flow by setting dynamic trading fees. We characterise the equilibrium via a coupled system of partial differential equations and derive tractable, approximate closed-form expressions for the equilibrium fees. Our analysis shows that the two-regime structure found in monopoly models persists under competition: pools alternate between raising fees to deter arbitrage and lowering fees to attract noise trading and increase volatility. Under competition, however, the switching boundary shifts from the oracle price to a weighted average of the oracle and the competitors' exchange rates. Our numerical experiments show that, holding total liquidity fixed, an increase in the number of competing DEXs reduces execution slippage for strategic liquidity takers and lowers fee revenue per DEX. Finally, the effect on noise traders’ slippage depends on market activity: they are worse off in low-activity markets but better off in high-activity ones.
    
\vspace{0.7em}
\noindent \emph{Keywords:} decentralized finance, automated market makers, optimal fees, arbitrageurs, noise trading.
\vspace{0.5em}
\newline
\emph{Mathematics Subject Classification (2020):} 93E20, 
91B70. 

\vspace{0.5em}
\noindent \emph{JEL Classification:} C61, 
G23,	
D53. 
\end{abstract}

\section{Introduction}

Automated market makers (AMMs) are the dominant trading mechanism within decentralised exchanges (DEXs). As of February 2026, AMM-based DEXs process approximately \textdollar6~billion in daily trading volume, accounting for roughly 6.5\% of the total DeFi traded volume.\footnote{CoinGecko, ``Decentralized Exchanges (DEX) by Trading Volume,'' \url{https://www.coingecko.com/en/exchanges/decentralized} (accessed 20 February 2026).} In an AMM, liquidity providers are compensated through trading fees paid by liquidity takers. Trading fees play a central role in AMM design and performance. Work such \cite{milionis2024automated, milionis2022automated, milionis2022quantifying} and \cite{bichuch2025price} show that fees are required to compensate liquidity providers for adverse-selection costs such as loss-versus-rebalancing (LVR), i.e., the systematic losses incurred by liquidity providers when arbitrageurs trade against stale pool prices. Moreover, as shown in \cite{hasbrouck2022need}, higher trading fees increase total trading volume by incentivizing liquidity provision, which in turn reduces price impact and makes the venue more attractive to liquidity takers.
\newline \newline
Given the foundational role of trading fees as a core building block of AMMs, it is a natural question to ask: what fee level is optimal from the perspective of a liquidity provider? This question is well posed in light of recent innovations that make fees a configurable choice rather than an exogenous parameter. In particular, with the introduction of Uniswap v4, liquidity providers can deploy and customise their own pools and, through \emph{hooks}, they are able to specify fee designs directly at the smart-contract level, allowing fee schedules that are dynamic and depend on external information, such as the asset’s price on a centralised exchange (e.g., Binance) or the asset price in another liquidity pool.
\newline \newline
There is a growing literature that studies optimal fee setting in automated market makers from the liquidity provider’s perspective. \cite{evans2021optimalfees} studies optimal fees in geometric mean market makers. \cite{fritsch2021note} determines optimal fees across multiple pools that compete with each other and relates the fee level to the depth of the pool. \cite{he2024optimaldesignautomatedmarket} derives an optimal fee for a liquidity provider who decides how to allocate their liquidity between a DEX and a centralised exchange (CEX). \footnote{There are numerous treatments of the CEX-DEX trading problem from the liquidity taker’s perspective; see \cite{cartea2023execution, jaimungal2023optimal, he2025arbitrage, bergault2026trading, capponi2026optimal}. For the liquidity provider’s perspective see \cite{bayraktar2024dex} and \cite{drissi2025equilibrium}.} All of these contributions differ from this work in that they consider static fees, whereas here we study the case of dynamic fees.
\newline \newline
Recent work argues that fee schedules should be state-dependent rather than fixed. \cite{cao2023structural} develops and estimates a structural model of an AMM and shows that fixed fees are inefficient. \cite{nadkarni2024adaptive} and \cite{cartea2024strategic} propose to adapt AMM bonding curves (which is equivalent to setting dynamic fees) deriving a range of an optimal adaptive curves that minimizes arbitrage losses while remaining competitive. \cite{campbell2025optimal} studies LP profitability in a dynamic model with a parallel CEX and endogenous order routing, and characterizes the optimal fee as a function of market conditions such as volatility and trading volume, suggesting a threshold-type dynamic fee schedule that is stable in normal conditions and changes with volatility. Finally, \cite{baggiani2025optimal} determines optimal dynamic fees in a constant function market maker. They find two distinct regimes (higher fees to deter arbitrageurs and lower fees to attract noise traders), and show that fee rules that are linear in inventory and sensitive to external price changes provide a good approximation to the optimal fee structure. 
\newline \newline
In practice, however, liquidity is not monopolised by a single pool. The same pair is typically traded across multiple pools (often across multiple fee tiers or venues), and liquidity takers can route orders to the pool offering the most attractive execution. This introduces a strategic interaction: each pool’s fee decision affects not only its own quoted prices and inventory dynamics, but also the dynamics of the order flow across competing pools.
\newline \newline
In this paper, we build on \cite{baggiani2025optimal} to extend the market-making framework of \cite{avellaneda2008high}\footnote{For broader background on market making models we refer to the books of \cite{cartea2015algorithmic} and \cite{gueant2016financial}.} to a multi-agent setting. The goal is that fee setting becomes a competitive interaction with multiple liquidity providers rather than a single-agent control problem. There are a number of papers that are close to this work within a limit order book (LOB) setup. \cite{boyce2025market} considers a reference market maker facing a parametrized competition and \cite{chilenje2025market} extend their analysis to find the Stackelberg equilibrium. \cite{baldacci2023mean} study a mean field of competing market makers against strategic takers, \cite{guo2025macroscopic} develops a stochastic-game formulation of macroscopic market making with price competition and, \cite{luo2021dynamic} models market making with price competition and incomplete information as a nonzero-sum stochastic differential game.
\newline \newline
To do this we build a finite-horizon model of fee competition between constant-function market makers (CFMMs) trading the same two assets. Each pool $i$ chooses two predictable fee processes: a fee for selling the risky asset $\{\mathfrak{p}^{i}_{t}\}_{t \in [0,T]}$, and a fee for buying it $\{\mathfrak{m}^{i}_{t}\}_{t \in [0,T]}$; these processes are stochastic and, at any time $t$, they may depend on all available information just before time $t$. In contrast to the monopoly setting, the feedback representation of optimal fees will no longer be one-dimensional objects because we expect them to depend both on the agent and the competitors inventories (and hence quoted exchange rates). For the case of two venues $\{a,b\}$ order flow to venue $i \in \{a,b\}$ is driven by two types of traders.
Arbitrageurs become more active when venue $i$'s effective quote (with fees) is misaligned with the CEX price and with the other venue's quote. At the same time, noise traders generate a baseline stream of orders that is largely
unrelated to price dislocations. This structure aims to capture the increased order flow due to CEX--DEX arbitrage and pool mispricing. Each pool maximises expected cumulative fee revenue taking the competitors’ fee policies as given. We characterise an approximate Nash equilibrium via a coupled system of dynamic programming equations on the joint inventory grid and derive equilibrium fee formulae as a generalization of the structure obtained in the monopoly problem.
\newline \newline
Our analysis delivers three main insights. First, the \emph{two-regime} structure of optimal fees persists under competition: pools still alternate between deterring arbitrage and stimulating noise trading. The key difference is that the switching boundary is no longer pinned to the oracle price alone; instead, it shifts to a weighted average of the oracle price and competitors’ exchange rates, reflecting that the relevant “outside option” for a liquidity taker is a combination of external execution and execution in rival pools. Second, equilibrium fees exhibit an approximately linear dependence on both own and rivals’ inventories over economically relevant regions, suggesting that a linear approximation of the fees (which lower the gas cost) remain effective even in multi-venue environments. Third, competition has welfare implications: as the number of competing venues increases, the ability of strategic liquidity takers to route to the best quotes improves, while fee revenues per venue decline and can become too small to justify entry when fixed participation costs are present.
\newline \newline
The remainder of the paper is organised as follows. Section \ref{section: two players model} introduces the two-player model, defines fee-dependent quotes, and specifies the controlled order-flow intensities. Section \ref{section: two player models numerical results} presents numerical equilibrium fee schedules and comparative statics in the duopoly. Section \ref{section: two player model simulations} studies execution outcomes and revenue comparisons across fee rules and against the monopoly benchmark. For readability the main body develops the mathematics for two players and Appendix \ref{section: M-player model} extends the framework to $M$ competing players. Lastly, Appendix \ref{section: M-player model simulations} reports the corresponding results for the multi-player case.

\section{Two-player model}\label{section: two players model}

We study the problem of two pools competing with each other in a finite horizon $T>0$. We assume that every pool is a CFMM and trades the same two assets: a riskless asset $X$ and a risky asset $Y$. We will denote the two players (pools) as player $a$ and player $b$. For $i \in \{a,b\}$, let $(\depth^{i})^{2}$ denote the depth of pool $i$ and let $f^{i}: \mathbb{R}_{+} \times \mathbb{R}_{+} \to \mathbb{R}_{+}$ be its trading function, with associated level function $\varphi^{i}: \mathbb{R}_{+} \to \mathbb{R}_{+}$ defined such that $f^{i}(\varphi^{i}(y),y) = (\depth^{i})^{2}$. Here, all trading functions are strictly increasing and twice differentiable in both arguments. For each pool we assume that the quantity of asset $Y$ (and consequently of asset $X$) takes values in a finite grid given by \begin{align} \label{eq: grid for asset Y}
    \{ y^{i,-N^{i}}, \dots, y^{i,0}, \dots, y^{i, N^{i}} \} \quad \quad i \in \{a,b\}.
\end{align}
Let $\underline{y}^{i} : =  y^{i,-N^{i}}$ and $\overline{y}^{i} : = y^{i,N^{i}}$ be the reserve constraints such that $0 < \underline{y}^{i} < y^{i,0} < \overline{y}^{i} < + \infty$.
Consequently, the quantity of asset $X$ in the pools takes values in the grid 
\begin{equation*}\label{eq: grid for quantity of x}
        \big\{ x^{i,-N^{i}} :=\varphi^{i}(y^{i,-N^{i}}), \dots, x^{i,0} := \varphi^{i}(y^{i,0}), \dots, x^{i,N^{i}} := \varphi^{i}(y^{i,N^{i}}) \big\}  \quad \quad i \in \{a,b\}.
\end{equation*}
Note that the monotonicity of the grids for the two assets are opposite to each other: the grid for asset $Y$ is increasing, whereas the grid for asset $X$ is decreasing.
The notation $y^{i,j}$ refers to the $j$-th element of player $i$'s grid.

For each pool there are three basic exchange rates to be considered. Let $y^{i,j}$ denote the quantity of asset $Y$ in the pool $i \in \{a,b\}$ before a trade of a liquidity taker, and let $\dot{\varphi}$ denote the derivative of $\varphi$. We define the following exchange rates.

\begin{enumerate}

    \item The \emph{marginal exchange rate} describes the price of an infinitesimal trade and is given by \begin{equation*}\label{eq: marginal exchange}
        Z^{i} ( y^{i,j} ) : = - \dot{\varphi}^{i} (y^{i,j}) \qquad j \in \{- N^{i}, \dots, N^{i} \}.
    \end{equation*}

    \item The \emph{exchange rate for buying} (taking out asset $Y$ from the pool) $\Delta^{i}_{-}(y^{i,j}):= y^{i,j} - y^{i,j-1}$ units of asset $Y$ with $j \in \{- N^{i}, \dots, N^{i} \}$ is given by
    \begin{equation*}\label{eq: exchange rate for buying}
        \exratebuy^{i}(y^{i,j}) :  = \frac{\varphi^{i}(y^{i,j-1}) - \varphi^{i}(y^{i,j})}{\Delta^{i}_{-}(y^{i,j})}.
    \end{equation*}
The quantity $\exratebuy^{i}(y^{i,j})$ takes values in the grid
    \begin{equation*}\label{eq: grid for buying exchange}
\left\{
  \exratebuy^{i,-N^{i} + 1} := \exratebuy^{i}(y^{i,-N^{i}+1}),
  \dots,
  \exratebuy^{i,0} := \exratebuy^{i}(y^{i,0}),
  \dots,
  \exratebuy^{i,N^{i}} := \exratebuy^{i}(y^{i,N^{i}})
\right\}.
    \end{equation*}
    
    \item The \emph{exchange rate for selling} (depositing asset $Y$ in the pool) $\Delta^{i}_{+}(y^{i,j}):= y^{i,j} - y^{i,j+1}$ units of asset $Y$ with $j \in \{- N^{i}, \dots, N^{i} \}$ is given by
    \begin{equation*}\label{eq: exchange rate for selling}
        \exratesell^{i}(y^{i,j}) :  = \frac{\varphi^{i}(y^{i,j}) - \varphi^{i}(y^{i,j+1})}{\Delta^{i}_{+}(y^{i,j})}.
    \end{equation*}

The quantity $\exratesell^{i}(y^{i,j})$ takes values in  the grid
    \begin{equation*}\label{eq: grid for selling exchange}
      \left\{ \exratesell^{i,-N^{i}} := \exratesell^{i}(y^{i,-N^{i}}), \dots, \exratesell^{i,0} := \exratesell^{i}(y^{i,0}), \dots, \exratesell^{i,N^{i} - 1} :=  \exratesell^{i}(y^{i,N^{i}-1}) \right\}.
    \end{equation*}
\end{enumerate}
For each player, exchange rates for buying and selling satisfy the identity 
\begin{equation*}
\exratebuy^{i,j} = \exratesell^{i,j-1}, \quad j \in \{ -N^{i} + 1, \dots, 0, \dots N^{i}\}, \quad \quad \quad i \in \{a,b\}.
\end{equation*}

The quantities $Z_{+}(y)$ and $Z_{-}(y)$ represent the bid and ask prices in the limit order book, respectively, while $Z(y)$ denotes the midprice. Thus, we have the ordering
\[
Z_{+}(y) < Z(y) < Z_{-}(y).
\]
Moreover, as the spread $\Delta(y)$ tends to zero,
\[
\lim_{\Delta(y)\to 0} Z_{-}(y) = \lim_{\Delta(y)\to 0} Z_{+}(y) = Z(y).
\]

We add fees to the trading mechanism, using the functions \[ \feeLTbuys^{i}, \feeLTsells^{i}:  \{ y^{i,-N^{i}}, \dots, y^{i,0}, \dots, y^{i,N^{i}} \} \times  \{ y^{j,-N^{j}}, \dots, y^{j,0}, \dots, y^{j,N^{j}}\} \to \R. \] Here, $\feeLTbuys$ denotes the fee for buying while $\feeLTsells$ the one for selling.
The fees are collected in units of the riskless asset $X$ and are deposited outside of the pool. The fees depend on inventory and, consequently, on the exchange rates quoted by both players. This contrasts with the monopoly case, where fees are one-dimensional. The associated (fee-dependent) exchange rates are therefore two-dimensional and are described as follows. Let $y^{i,h}$ and $y^{j,s}$ denote the quantities of asset $Y$ in pools $i$ and $j$, respectively, immediately before a trade.
 \begin{enumerate}
    \item If $h-1 \geq -N^{i}$, the \emph{exchange rate with fee structure $\feeLTbuys^{i}$ for buying} (taking out asset $Y$ from the pool) $\Delta^{i}_{-}(y^{i,h}):= y^{i,h}-y^{i,h-1}$ units of asset $Y$ is given by
    \begin{equation*}\label{eq: exchange rate for buying with fees}
        \exratebuy^{i,\feeLTbuys^{i}}(y^{i,h}) : = (1 + \feeLTbuys^{i}(y^{i,h},y^{j,s})) \exratebuy^{i}(y^{i,h}).
    \end{equation*}
    \item If $h+1 \leq N^{i}$, the \emph{exchange rate with fee structure $\feeLTsells^{i}$ for selling} (depositing asset $Y$ in the pool) $\Delta^{i}_{+}(y^{i,h}):= y^{i,h+1}-y^{i,h}$ units of asset $Y$ is given by
    \begin{equation*}\label{eq: exchange rate for selling with fees}
        \exratesell^{i,\feeLTsells^{i}}(y^{i,h}) : = (1 - \feeLTsells^{i}(y^{i,h},y^{j,s})) \exratesell^{i}(y^{i,h}).
    \end{equation*}
\end{enumerate}
The AMM’s mechanism is independent of the fee schedule: the fee revenues are routed to an external cash account and redistributed later. 

Our model aims to study how different pools compete in setting fees dynamically to maximize revenue coming from the order flow. We work in a probability space $(\Omega, \mathcal{F}, \mathbb{F} = \{ \mathcal{F}_{t} \}_{t \in [0,T]}, \mathbb{P}^{\feeLTsells,\feeLTbuys})$ supporting all the processes we use below. We model one representative liquidity taker (LT) and assume that their buy and sell order arrivals to each pool $i \in \{a,b\}$ are described by point processes $\{ N^{i,-,\feeLTbuys_{t}} \}_{t \in [0,T]}$ and $\{  N^{i,+,\feeLTsells_{t}} \}_{t \in [0,T]}$ where $\feeLTbuys := (\feeLTbuys^{a},\feeLTbuys^{b})$, $\feeLTsells := (\feeLTsells^{a},\feeLTsells^{b})$ and $\feeLTbuys^{i}$ and $\feeLTsells^{i}$ are assumed to be predictable for each $i \in \{a,b\}$. We include the fees processes $\mathfrak{p}_{t}^{i}$ and $\mathfrak{m}_{t}^{i}$ in the subscript of $\{ N^{i,-,\feeLTbuys_{t}} \}_{t \in [0,T]}$ and $\{  N^{i,+,\feeLTsells_{t}} \}_{t \in [0,T]}$ to draw attention to the controlled stochastic intensities under the probability measure $\mathbb{P}^{\feeLTsells,\feeLTbuys}$. In this work, the depths of the pools $\depth^{i}$ are fixed throughout $[0,T]$, that is, $T$ is small enough so that LPs do not add or remove liquidity from the pool. This allows us to study liquidity taking and optimal fees in isolation.
This assumption can be relaxed by allowing the total depth of each pool to evolve on a fixed grid over $[0,T]$. In that setting, changes in liquidity may arise either exogenously, due to noise liquidity providers, or endogenously, from strategic liquidity providers responding to market conditions. We leave the analysis of both extensions for future research.

Next, we model the intensity of liquidity taking orders arriving to each pool. Let us denote by $\mathcal{A}$ the space of $\mathbb{F}$-predictable and bounded processes and let $\fees^{i} : = ( \feeLTbuys^{i}, \feeLTsells^{i}) \in \mathcal{A}$ for $i \in \{a,b\}$. The quantity of asset $Y$ in the pool $i$ at time $t \in [0,T]$ is given by 
\[ Y_{t}^{i,\fees} : = y^{N^{i,+,\feeLTsells}_{t}  -  N^{i,+,\feeLTbuys}_{t}}, \]
where $\fees := (\fees^{a}, \fees^{b})$.
For each $i \in \{a,b\}$ and $j \in \{a,b\}$, $j \neq i$ the controlled intensities of $\{ N^{i,-,\feeLTbuys_{t}} \}_{t \in [0,T]}$ and $\{  N^{i,+,\feeLTsells_{t} } \}_{t \in [0,T]}$ are given by

\begin{align*}
\lambda_{t}^{i,-,\fees} & := \lambda^{i,-} \exp{ \left( k^{i,0}((S_{t} - \zeta)  - \exratebuy^{i,\feeLTbuys^{i}}(Y_{t}^{i,\fees}))\Delta_{-}^{i}(Y_{t}^{i,\fees}) + k^{i,j}(\exratebuy^{j}(Y_{t}^{j,\fees}) - \exratebuy^{i,\feeLTbuys^{i}}(Y_{t}^{i,\fees})) \Delta_{-}^{i}(Y_{t}^{i,\fees}) \right) } \ind_{ \{Y_{t}^{i,\fees} > \underline{y}^{i} \}}, \\
\lambda_{t}^{i,+,\fees} & := \lambda^{i,+} \exp{ \left( k^{i,0}(\exratesell^{i,\feeLTsells^{i}}(Y_{t}^{i,\fees}) - (S_{t} + \zeta)) \Delta_{+}^{i}(Y_{t}^{i,\fees}) + k^{i,j}(\exratesell^{i,\feeLTsells^{i}}(Y_{t}^{i,\fees}) - \exratesell^{j}(Y_{t}^{j,\fees})) \Delta_{+}^{i}(Y_{t}^{i,\fees}) \right)  } \ind_{ \{Y_{t}^{i,\fees} < \overline{y}^{i} \}}.
\end{align*}

Here, $\lambda^{i,-}$ and $\lambda^{i,+}$ are some baseline intensities, $k^{i,0}$ and $k^{i,j}$ are positive exponential decay parameters,\footnote{Note that different pools get different parameters because there could be external factors (depth) that influence the trading flows in the pool.} $\{\oraclepricestochastic\}_{t \in [0, T]}$ denotes the midprice in a centralised exchange (outside of the pool) of asset $Y$ in terms of the riskless asset $X$, the so-called \emph{oracle price}, and $\zeta > 0$ is the corresponding half-spread in the centralised exchange. For simplicity, $\{\oraclepricestochastic\}_{t \in [0, T]}$ has dynamics given by \[ \oraclepricestochastic = S_0 + \sigma W_{t}, \] where $\{ W_{t} \}_{t \in [0,T]}$ is a Brownian motion. Notice that we recover the model from the monopoly case studied in \cite{baggiani2025optimal} by setting the parameters $k^{i,j} = 0$.

The controlled stochastic intensities $\lambda_{t}^{i,-,\fees}$ and $\lambda_{t}^{i,+,\fees}$ model that when $\exratebuy^{i,\feeLTbuys^{i}_{t}}(Y_{t-}^{\fees}) < \oraclepricestochastic - \zeta$, buying $\Delta^{i}_{-}(y^{i,j})$ units to the pool and selling the same amount from the external venue is profitable to arbitrageurs, in a similar way when $\exratebuy^{i,\feeLTbuys^{i}_{t}}(Y_{t-}^{i,\fees}) < \exratesell^{j}(Y_{t-}^{j,\fees})$ arbitrageurs can do a roundtrip trade between the two pools. These two effects make the buy intensity increase. There is a similar effect on the sell intensity. In order to make the notation simpler we set $\zeta =0$. The mathematical results can all be obtained for $\zeta > 0$.

\medskip{}
For each $i \in \{a,b\}$ the cumulative fees $\{\Cash_{t}^{i,\fees^{i}}\}_{t \in [0, T]}$ collected by the pools are in turn given by
\begin{equation*} \label{eq : SDE for holdings}
\Cash_{t}^{i,\fees^{i}} := \int_0^t \feeLTsells_{u}^{i} \exratesell^{i,\feeLTsells_{u}^{i}}(Y_{u}^{i,\fees}) \Delta_{+}^{i}(Y_{t}^{i,\fees}) \dd N^{i,+,\feeLTsells^{i}}_{u} + \int_0^t \feeLTbuys^{i}_{u}\exratebuy^{i,\feeLTbuys^{i}_{u}}(Y_{u}^{i,\fees}) \Delta_{-}^{i}(Y_{t}^{i,\fees}) \dd N^{i,+,\feeLTbuys^{i}}_{u}, \quad \quad t \in [0,T].
\end{equation*}
The LPs seek to solve the control problems
\begin{align*}
    v^{a}(t,s,\cash^{a},y^{a},y^{b}) & : = \sup_{(\feeLTbuys^{a},\feeLTsells^{a}) \in \mathcal{A}_{t}} v^{(a,\feeLTbuys,\feeLTsells)}(t,s,\cash,y^{a},y^{b}), \\
    v^{b}(t,s,\cash^{b},y^{a},y^{b}) & : = \sup_{(\feeLTbuys^{b},\feeLTsells^{b}) \in \mathcal{A}_{t}} v^{(b,\feeLTbuys,\feeLTsells)}(t,s,\cash,y^{a},y^{b}),
\end{align*}
where $\mathcal{A}_{t}$ denotes the set of all $\mathbb{F}$-predictable and bounded fee structure processes $(\feeLTbuys_{u}^{i},\feeLTsells_{u}^{i})_{\{t \leq u \leq T\}}$ and the conditional performance criterion are given by \[ v^{(a,\feeLTbuys,\feeLTsells)}(t,s,y^{a},y^{b}) : = \mathbb{E}_{(t,s,\cash^{a},y^{a},y^{b})}^{(\feeLTbuys,\feeLTsells)} \left[ \Cash_{T}^{(a,t,s,y^{a},y^{b},\fees^{a})} \right], \quad \quad  v^{(b,\feeLTbuys,\feeLTsells)}(t,s,y^{a},y^{b}) : = \mathbb{E}_{(t,s,\cash^{b},y^{a},y^{b})}^{(\feeLTbuys,\feeLTsells)} \left[ \Cash_{T}^{(b,t,s,y^{a},y^{b},\fees^{b})} \right]. \]
Here, \[ \{ Y_{u}^{(a,t,s,\cash^{a},y^{a},y^{b},\fees^{a})} \}_{u \in [t,T]}, \; \{ Y_{u}^{(b,t,s,\cash^{b},y^{a},y^{b},\fees^{b})} \}_{u \in [t,T]} \text{ and } \{ S_{u}^{(t,s,\cash^{a},\cash^{b},y^{a},y^{b})}\}_{u \in [t,T]} \] denote the (controlled) processes $Y^{a}$, $Y^{b}$ and $S$ restarted at time $t$ with initial value $\cash^{a}$, $\cash^{b}$, $y^{a}$, $y^{b}$ and $s$, respectively. From the dynamic programming principle, we determine that the Hamilton-Jacobi-Bellman (HJB) for the value function of the player $a$ is
\small
\begin{equation*}
\begin{aligned}
0 ={}\;& \partial_t v^{a}(t,s,y^{a},y^{b})
      + \frac{\sigma^{2}}{2}\,\partial_{ss} v^{a}(t,s,y^{a},y^{b}) \\
& + \sup_{\feeLTbuys^{a}\in\mathbb{R}}
\Bigg[
\lambda^{a,-}
\exp\!\Big(
k^{a}\Big(
\frac{k^{a,0}}{k^{a}}\,s
+ \frac{k^{a,b}}{k^{a}}\,\exratebuy^{b}(y^{b})
- (1+\feeLTbuys^{a})\,\exratebuy^{a}(y^{a})
\Big)\Delta_{-}^{a}(y^{a})
\Big) \\
&\qquad\qquad\qquad\qquad\qquad \times
\Big(
v^{a}(t,s,y^{a}-\Delta_{-}^{a}(y^{a}),y^{b})
- v^{a}(t,s,y^{a},y^{b})
+ \feeLTbuys^{a}\,\exratebuy^{a}(y^{a})\,\Delta_{-}^{a}(y^{a})
\Big)\,\ind_{\{y^{a}>\underline{y}^{a}\}}
\Bigg] \\
& + \lambda^{b,-}
\exp\!\Big(
k^{b}\Big(
\frac{k^{b,0}}{k^{b}}\,s
+ \frac{k^{b,a}}{k^{b}}\,\exratebuy^{a}(y^{a})
- (1+\feeLTbuys^{b})\,\exratebuy^{b}(y^{b})
\Big)\Delta_{-}^{b}(y^{b})
\Big) \\
&\qquad\qquad\qquad\qquad\qquad \times
\Big(
v^{a}(t,s,y^{a},y^{b}-\Delta_{-}^{b}(y^{b}))
- v^{a}(t,s,y^{a},y^{b})
\Big)\,\ind_{\{y^{b}>\underline{y}^{b}\}} \\
& + \sup_{\feeLTsells^{a}\in\mathbb{R}}
\Bigg[
\lambda^{a,+}
\exp\!\Big(
k^{a}\Big(
(1-\feeLTsells^{a})\,\exratesell^{a}(y^{a})
- \frac{k^{a,0}}{k^{a}}\,s
- \frac{k^{a,b}}{k^{a}}\,\exratesell^{b}(y^{b})
\Big)\Delta_{+}^{a}(y^{a})
\Big) \\
&\qquad\qquad\qquad\qquad\qquad \times
\Big(
v^{a}(t,s,y^{a}+\Delta_{+}^{a}(y^{a}),y^{b})
- v^{a}(t,s,y^{a},y^{b})
+ \feeLTsells^{a}\,\exratesell^{a}(y^{a})\,\Delta_{+}^{a}(y^{a})
\Big)\,\ind_{\{y^{a}<\overline{y}^{a}\}}
\Bigg] \\
& + \lambda^{b,+}
\exp\!\Big(
k^{b}\Big(
(1-\feeLTsells^{b})\,\exratesell^{b}(y^{b})
- \frac{k^{b,0}}{k^{b}}\,s
- \frac{k^{b,a}}{k^{b}}\,\exratesell^{a}(y^{a})
\Big)\Delta_{+}^{a}(y^{a})
\Big) \\
&\qquad\qquad\qquad\qquad\qquad \times
\Big(
v^{a}(t,s,y^{a},y^{b}+\Delta_{+}^{b}(y^{b}))
- v^{a}(t,s,y^{a},y^{b})
\Big)\,\ind_{\{y^{b}<\overline{y}^{b}\}} .
\end{aligned}
\end{equation*}
where $k^{a} = k^{a,0} + k^{a,b}$ and $k^{b} = k^{b,0} + k^{b,a}$. The HJB for player $b$ is similar and we omit it for brevity. First order conditions on the maximizers yields

\begin{align*}
    \feeLTbuys^{a,*}(t,s,y^{a},y^{b}) & = \frac{ 1 + k^{a}(v^{a}(t,s,y^{a},y^{b}) - v^{a}(t,s,y^{a} - \Delta_{-}^{a}(y^{a}),y^{b}))}{k^{a} \exratebuy^{a}(y^{a}) \Delta_{-}^{a}(y^{a})}, \\
    \feeLTsells^{a,*}(t,s,y^{a},y^{b}) & = \frac{ 1 + k^{a}(v^{a}(t,s,y^{a},y^{b}) - v^{a}(t,s,y^{a} + \Delta_{+}^{a}(y^{a}),y^{b}))}{k^{a} \exratesell^{a}(y^{a}) \Delta_{+}^{a}(y^{a})}, \\
    \feeLTbuys^{b,*}(t,s,y^{a},y^{b}) & = \frac{ 1 + k^{b}(v^{b}(t,s,y^{a},y^{b}) - v^{b}(t,s,y^{a},y^{b} - \Delta_{-}^{b}(y^{b})))}{k^{b} \exratebuy^{b}(y^{b}) \Delta_{-}^{b}(y^{b})}, \\
    \feeLTsells^{b,*}(t,s,y^{a},y^{b}) & = \frac{ 1 + k^{b}(v^{b}(t,s,y^{a},y^{b}) - v^{b}(t,s,y^{a},y^{b}  + \Delta_{+}^{b}(y^{b})))}{k^{b} \exratesell^{b}(y^{b}) \Delta_{+}^{b}(y^{b})}.
\end{align*}

Thus, the HJB for player $a$ turns into the following PDE
\begin{align*}
  0 = & \frac{\partial}{\partial t} v^{a}(t,s,y^{a},y^{b}) +  \frac{\sigma^{2}}{2} \frac{\partial^{2}}{\partial s^{2}} v^{a}(t,s,y^{a},y^{b}) \\
    & \frac{\lambda^{a,-} e^{ k^{a} \left( \frac{k^{a,0}}{k^{a}} s + \frac{k^{a,b}}{k^{a}} \exratebuy^{b}(y^{b}) \right) \Delta^{a}_{-}(y^{a}) - 1}}{k^{a}} e^{-k^{a} \exratebuy^{a}(y^{a}) \Delta_{-}^{a}(y^{a})} e^{k^{a}(v^{a}(t,s,y^{a} - \Delta_{-}^{a}(y^{a}),y^{b}) - v^{a}(t,s,y^{a},y^{b}))} \ind_{ \{ y^{a} > \underline{y}^{a} \} } \; + \\
    & \lambda^{b,-} e^{ k^{b} \left( \frac{k^{b,0}}{k^{b}} s + \frac{k^{b,a}}{k^{b}} \exratebuy^{a}(y^{a}) \right) \Delta^{b}_{-}(y^{b}) - 1} e^{-k^{b} \exratebuy^{b}(y^{b}) \Delta_{-}^{b}(y^{b})}\\ & \quad \quad \quad \quad \quad \quad \quad \quad \quad e^{k^{b} (v^{b}(t,s,y^{a}, y^{b} - \Delta_{-}^{b}(y^{b})) - v^{b}(t,s,y^{a}, y^{b}))} \left( v^a(t, s, y^a, y^b - \Delta_{-}^{b}(y^{b})) - v^a(t, s, y^a, y^b) \right) \ind_{ \{ y^{b} > \underline{y}^{b} \} } \; + \\
    & \frac{\lambda^{a,+} e^{ - k^{a} \left( \frac{k^{a,0}}{k^{a}} s + \frac{k^{a,b}}{k^{a}} \exratesell^{b}(y^{b}) \right) \Delta^{a}_{+}(y^{a}) - 1}}{k^{a}} e^{k^{a} \exratesell^{a}(y^{a}) \Delta_{+}^{a}(y^{a})} e^{k^{a}(v^{a}(t,s,y^{a} + \Delta_{+}^{a}(y^{a}),y^{b}) - v^{a}(t,s,y^{a},y^{b}))} \ind_{ \{ y^{a} < \overline{y}^{a} \} } \; + \\
    & \lambda^{b,+} e^{ - k^{b} \left( \frac{k^{b,0}}{k^{b}} s + \frac{k^{b,a}}{k^{b}} \exratesell^{a}(y^{a}) \right) \Delta^{b}_{+}(y^{b}) - 1} e^{k^{b} \exratesell^{b}(y^{b}) \Delta_{+}^{b}(y^{b})}\\ & \quad \quad \quad \quad \quad \quad \quad \quad \quad e^{k^{b} (v^{b}(t,s,y^{a}, y^{b} + \Delta_{+}^{b}(y^{b})) - v^{b}(t,s,y^{a}, y^{b}))} \left( v^a(t, s, y^a, y^b + \Delta_{+}^{b}(y^{b})) - v^a(t, s, y^a, y^b) \right) \ind_{ \{ y^{b} < \overline{y}^{b} \} }.
\end{align*}
We now make a few simplifying assumptions to compute a closed-form solution. More precisely, we treat $s$ as a parameter and we assume that the change in one agent's inventory has a negligible impact on the other player's value function, that is
\begin{align*}
     v^{a}(t,s,y^{a} ,y^{b} + \Delta_{+}^{b}(y^{b})) \approx v^{a}(t,s,y^{a},y^{b}), \qquad
     v^{a}(t,s,y^{a} ,y^{b} - \Delta_{-}^{b}(y^{b}))  \approx v^{a}(t,s,y^{a},y^{b}).
\end{align*}

\begin{remark}
    Our results indicate that the above approximation provides a good level of accuracy for the purposes of this study, particularly in good regimes where the within-venue price $Z$ remains close to the centralised reference price $S$.\footnote{See the project repository at \url{https://github.com/leonardobaggiani/amm-fees-competition} for the implementation details.}
\end{remark}
The approximated equation takes the form of

\begin{align*}
     0 = & \frac{\partial}{\partial t} v^{a}(t,y^{a},y^{b}) \\
    & + \frac{\lambda^{a,-} e^{ k^{a} \left( \frac{k^{a,0}}{k^{a}} s + \frac{k^{a,b}}{k^{a}} \exratebuy^{b}(y^{b}) \right) \Delta^{a}_{-}(y^{a}) - 1}}{k^{a}} e^{-k^{a} \exratebuy^{a}(y^{a}) \Delta_{-}^{a}(y^{a})} e^{k^{a}(v^{a}(t,y^{a} - \Delta_{-}^{a}(y^{a}),y^{b}) - v^{a}(t,y^{a},y^{b}))} \ind_{ \{ y^{a} > \underline{y}^{a} \} } \; \\
    & + \frac{\lambda^{a,+} e^{ - k^{a} \left( \frac{k^{a,0}}{k^{a}} s + \frac{k^{a,b}}{k^{a}} \exratesell^{b}(y^{b}) \right) \Delta^{a}_{+}(y^{a}) - 1}}{k^{a}} e^{k^{a} \exratesell^{a}(y^{a}) \Delta_{+}^{a}(y^{a})} e^{k^{a}(v^{a}(t,y^{a} + \Delta_{+}^{a}(y^{a}),y^{b}) - v^{a}(t,y^{a},y^{b}))} \ind_{ \{ y^{a} < \overline{y}^{a} \} }.
\end{align*}
We make the ansatzes

\begin{align*}
    e^{k^{a}v^{a}(t,y^{a},y^{b})}  : = w^{a}(t,y^{a},y^{b}), \qquad
    e^{k^{b}v^{b}(t,y^{a},y^{b})}  : = w^{b}(t,y^{a},y^{b}),
\end{align*}
and we get
\begin{align*}
    0 = \frac{\partial}{\partial t} w^{a}(t,y^{a},y^{b}) & + \lambda^{a,-} e^{ k^{a} \left( \frac{k^{a,0}}{k^{a}} s + \frac{k^{a,b}}{k^{a}} \exratebuy^{b}(y^{b}) \right) \Delta^{a}_{-}(y^{a}) - 1} e^{-k^{a} \exratebuy^{a}(y^{a}) \Delta_{-}^{a}(y^{a})} w^{a}(t,y^{a} - \Delta_{-}^{a}(y^{a}),y^{b}) \ind_{ \{ y^{a} > \underline{y}^{a} \} } \; \\
    & + \lambda^{a,+} e^{ - k^{a} \left( \frac{k^{a,0}}{k^{a}} s + \frac{k^{a,b}}{k^{a}} \exratesell^{b}(y^{b}) \right) \Delta^{a}_{+}(y^{a}) - 1} e^{k^{a} \exratesell^{a}(y^{a}) \Delta_{+}^{a}(y^{a})} w^{a}(t,y^{a} + \Delta_{-}^{a}(y^{a}),y^{b}) \ind_{ \{ y^{a} < \overline{y}^{a} \} }.
\end{align*}
An explicit solution of the above PDEs is given in the following theorem.

\begin{theorem} \label{th: nash equilibrium theorem}
    For $l \in \{-N^{b}, \dots , N^{b} \}$ and $h \in \{-N^{a}, \dots , N^{a} \}$ define the matrices $ \mathbf{A}^{b,l} : =  ( \mathbf{A}^{l}_{i,j} )_{0 \leq i \leq j \leq 2N}$ and $ \mathbf{A}^{a} : =  ( \mathbf{A}^{a,h}_{i,l} )_{0 \leq i \leq l \leq 2N}$ by 
    \begin{align*}
    \mathbf{A}^{a}_{i,j}(y^{b,l}) & : =
    \begin{cases}
        \lambda^{a,+}e^{- k^{a} \left( \frac{k^{a,0}}{k^{a}} s + \frac{k^{a,b}}{k^{a}} \exratesell^{b}(y^{b,l}) \right) \Delta^{a}_{+}(y^{a,j-N^{a}}) - 1} e^{k^{a} \exratesell^{a}(y^{a,j-N^{a}}) \Delta_{+}^{a}(y^{a,j-N^{a}})} & \text{ if } i = j - 1, \\
        \lambda^{a,-} e^{ k^{a} \left( \frac{k^{a,0}}{k^{a}} s + \frac{k^{a,b}}{k^{a}} \exratebuy^{b}(y^{b,l}) \right) \Delta^{a}_{-}(y^{a,j-N^{a}}) - 1} e^{-k^{a} \exratebuy^{a}(y^{a,j-N^{a}}) \Delta_{-}^{a}(y^{a,j-N^{a}})} & \text{ if } i = j + 1 \\
        0 & \text{ otherwise,}
    \end{cases} \\
    \mathbf{A}^{b}_{i,j}(y^{a,h}) & : =
    \begin{cases}
        \lambda^{b,+}e^{- k^{b} \left( \frac{k^{b,0}}{k^{b}} s + \frac{k^{b,a}}{k^{b}} \exratesell^{a}(y^{a,h}) \right) \Delta^{b}_{+}(y^{b,j-N^{b}}) - 1} e^{k^{b} \exratesell^{b}(y^{b,j-N^{b}}) \Delta_{+}^{b}(y^{b,j-N^{b}})} & \text{ if } i = j - 1, \\
        \lambda^{b,-} e^{ k^{b} \left( \frac{k^{b,0}}{k^{b}} s + \frac{k^{b,a}}{k^{b}} \exratebuy^{a}(y^{a,h}) \right) \Delta^{b}_{-}(y^{b,j-N^{b}}) - 1} e^{-k^{b} \exratebuy^{b}(y^{b,j-N^{b}}) \Delta_{-}^{b}(y^{b,j-N^{b}})} & \text{ if } i = j + 1 \\
        0 & \text{ otherwise.}
    \end{cases}
\end{align*}
Denote with $\mathbf{1}$ the unit vectors of $\mathbb{R}^{2N^{a} + 1}$ and $\mathbb{R}^{2N^{b} + 1}$. Define the functions $w^{a}: [0,T] \times \{ y^{a,-N^{a}}, \dots, y^{a,N^{a}} \} \times \{ y^{b,-N^{b}}, \dots, y^{b,N^{b}} \} \to \mathbb{R}$ and $w^{b}: [0,T] \times \{ y^{a,-N^{a}}, \dots, y^{a,N^{a}} \} \times \{ y^{b,-N^{b}}, \dots, y^{b,N^{b}} \} \to \mathbb{R}$ by 
\begin{align*}
    w^{a}(t,y^{a,i},y^{b,j}) &:= \left( \exp\big( \mathbf{A}^{a}(y^{b,j})(T - t) \big) \, \mathbf{1} \right)_{i} \\
    w^{b}(t,y^{a,i},y^{b,j}) &:= \left( \exp\big( \mathbf{A}^{b}(y^{a,i})(T - t) \big) \, \mathbf{1} \right)_{j}
\end{align*}
and the functions $v^{a}: [0,T] \times \{ y^{a,-N^{a}}, \dots, y^{a,N^{a}} \} \times \{ y^{b,-N^{b}}, \dots, y^{b,N^{b}} \} \times \mathbb{R}_{+} \to \mathbb{R}$, $v^{b}: [0,T] \times \{ y^{a,-N^{a}}, \dots, y^{a,N^{a}} \} \times \{ y^{b,-N^{b}}, \dots, y^{b,N^{b}} \} \times \mathbb{R}_{+} \to \mathbb{R}$ as 
\begin{align*}
    v^{a}(t,y^{a,i},y^{b,j}, \cash^{a}) & := \cash^{a} + \frac{1}{k^{a}} \log(w^{a}(t,y^{a,i},y^{b,j})), \\
    v^{b}(t,y^{a,i},y^{b,j}, \cash^{b}) & := \cash^{b} + \frac{1}{k^{b}} \log(w^{b}(t,y^{a,i},y^{b,j})). \\
\end{align*}
Then $v^{a}$ and $v^{b}$ solve the system of HJBs
\begin{align*}
\begin{cases}
0 \;=\;&
\frac{\partial}{\partial t}\, v^{a}(t,y^{a,i},y^{b,j}, \cash^{a}) \\[4pt]
&\;+\; \sup_{\feeLTbuys^{a}\in\mathbb{R}} \lambda^{a,-}\,
\exp\Bigg\{ k^{a}\!\Big( \tfrac{k^{a,0}}{k^{a}}\,s
+ \tfrac{k^{a,b}}{k^{a}}\,\exratebuy^{b}(y^{b}) \Big) \\
&\qquad\qquad
- k^{a}\!\Big(1+\feeLTbuys^{a}\Big)\exratebuy^{a}(y^{a}) \,\Delta_-^{a}(y^{a}) \Bigg\}
\Big( v^{a}(t,y^{a} - \Delta_{-}^{a}(y^{a}),y^{b}, \cash^{a} + \feeLTbuys^{a}\,\exratebuy^{a}(y^{a})\,\Delta_-^{a}(y^{a}))
      - v^{a}(t,y^{a},y^{b},\cash^{a}) \Big)\,
\ind_{\{y^{a}>\underline y^{a}\}} \\[6pt]
&\;+\; \sup_{\feeLTsells^{a}\in\mathbb{R}} \lambda^{a,+}\,
\exp\Bigg\{ -k^{a}\!\Big( \tfrac{k^{a,0}}{k^{a}}\,s
+ \tfrac{k^{a,b}}{k^{a}}\,\exratesell^{b}(y^{b}) \Big) \\
&\qquad\qquad
+ k^{a}\!\Big(1-\feeLTsells^{a}\Big)\exratesell^{a}(y^{a}) \,\Delta_+^{a}(y^{a}) \Bigg\}
\Big( v^{a}(t,y^{a} + \Delta_{+}^{a}(y^{a}),y^{b}, \cash^{a} + \feeLTsells^{a}\,\exratesell^{a}(y^{a})\,\Delta_+^{a}(y^{a}))
      - v^{a}(t,y^{a},y^{b},\cash^{a}) \Big)\,
\ind_{\{y^{a} < \overline y^{a}\}}, \\[6pt]
0 \;=\;&
\frac{\partial}{\partial t}\, v^{b}(t,s,y^{a},y^{b},\cash^{b}) \\[4pt]
&\;+\; \sup_{\feeLTbuys^{b}\in\mathbb{R}} \lambda^{b,-}\,
\exp\Bigg\{ k^{b}\!\Big( \tfrac{k^{b,0}}{k^{b}}\,s
+ \tfrac{k^{b,a}}{k^{b}}\,\exratebuy^{a}(y^{a}) \Big) \\
&\qquad\qquad
- k^{b}\!\Big(1+\feeLTbuys^{b}\Big)\exratebuy^{b}(y^{b}) \,\Delta_-^{b}(y^{b}) \Bigg\}
\Big( v^{b}(t,y^{a},y^{b}  - \Delta_{-}^{b}(y^{b}),\cash^{b} + \feeLTbuys^{b}\,\exratebuy^{b}(y^{b})\,\Delta_-^{b}(y^{b}))
      - v^{b}(t,y^{a},y^{b}, \cash^{b})
       \Big)\,
\ind_{\{y^{b}>\underline y^{b}\}} \\[6pt]
&\;+\; \sup_{\feeLTsells^{b}\in\mathbb{R}} \lambda^{b,+}\,
\exp\Bigg\{ -k^{b}\!\Big( \tfrac{k^{b,0}}{k^{b}}\,s
+ \tfrac{k^{b,a}}{k^{b}}\,\exratesell^{a}(y^{a}) \Big) \\
&\qquad\qquad
+ k^{b}\!\Big(1-\feeLTsells^{b}\Big)\exratesell^{b}(y^{b}) \,\Delta_+^{b}(y^{b}) \Bigg\}
\Big( v^{b}(t,y^{a},y^{b}  + \Delta_{+}^{b}(y^{b}), \cash^{b} + \feeLTsells^{b}\,\exratesell^{b}(y^{b})\,\Delta_+^{b}(y^{b}))
      - v^{b}(t,y^{a},y^{b}, \cash^{b})
      \Big)\,
\ind_{\{y^{b} < \overline y^{b}\}} \\[6pt]
\end{cases}
\end{align*}
with boundary conditions $v^{a}(T,y^{a},y^{b},\cash^{b}) = 0$ and $v^{b}(T,y^{a},y^{b},\cash^{b}) = 0$, for every $y^{a}$ and $y^{b}$. Moreover, the corresponding maximizers are independent of $\cash^{a}$ and $\cash^{b}$ and are given by
\begin{align*}
    \feeLTbuys^{a,*}(t,y^{a,i},y^{b,j}) & = \frac{1}{k^{a}\exratebuy^{a}(y^{a,i})\Delta_{-}^{a}(y^{a,i})} \left(1 + \log \left( \frac{w^{a}(t,y^{a,i},y^{b,j})}{w^{a}(t,y^{a,i-1},y^{b,j})} \right) \right), \\
    \feeLTsells^{a,*}(t,y^{a,i},y^{b,j}) & = \frac{1}{k^{a}\exratesell^{a}(y^{a,i})\Delta_{+}^{a}(y^{a,i})} \left(1 + \log \left( \frac{w^{a}(t,y^{a,i},y^{b,j})}{w^{a}(t,y^{a,i+1},y^{b,j})} \right) \right), \\
    \feeLTbuys^{b,*}(t,y^{a,i},y^{b,j}) & = \frac{1}{k^{b}\exratebuy^{b}(y^{b,j})\Delta_{-}^{b}(y^{b,j})} \left(1 + \log \left( \frac{w^{b}(t,y^{a,i},y^{b,j})}{w^{b}(t,y^{a,i},y^{b,j-1})} \right) \right), \\
    \feeLTsells^{b,*}(t,y^{a,i},y^{b,j}) & = \frac{1}{k^{b}\exratesell^{b}(y^{b,j})\Delta_{+}^{b}(y^{b,j})} \left(1 + \log \left( \frac{w^{b}(t,y^{a,i},y^{b,j})}{w^{b}(t,y^{a,i},y^{b,j+1})} \right) \right).
\end{align*}

\end{theorem}

\begin{remark} \label{rem: optimal fees connection with k}
    In \cite{avellaneda2008high}, the optimal strategy establishes a connection between $1/\kappa$ and the half spread in the market ($\kappa$ is a parameter in their model controlling the price sensitivity of the order flow). Similar to their paper, in our setup the quantity 
    $$
    1/\Big( k^a\,Z^{a,b}_{\pm} \Delta^{a,b}_{\pm}\Big)\,,
    $$
    establishes a connection between $k^a$ and the typical fee charged in the pool.
\end{remark}

\section{Numerical results}\label{section: two player models numerical results}

Here, we use constant product market makers, i.e., the trading functions $f^{i}: \mathbb{R}_{+} \times \mathbb{R}_{+} \to \mathbb{R}_{+}$ is $f^{i}(x,y) = x y$ and the level functions $\varphi: \mathbb{R}_{+} \to \mathbb{R}_{+}$ are $\varphi^{i}(y) = (p^{i})^{2} / y$, for every $i \in \{a,b\}$. Moreover, we set the parameters as \[ (p^{i})^2 = 10^{8} / 4, \quad \lambda^{i,+} = \lambda^{i,-} = 50, \quad k^{i,0} = k^{i,j} = 2,\] for every $i \in \{a,b\}$ and $j \in \{a,b\} \setminus \{i\}$. Moreover, we assume that the grids for the risky asset of two players are such that after every trade the exchange rate $Z^{i}(y)$ moves by $0.1$. Specifically, the grids are given by the formula \[ y^{i,j} : = \sqrt{\frac{(p^{i})^{2}}{Z^{i}(y^{i,0}) -0.1 \, j } }, \quad \quad \text{ for } i \in \{a,b\} \quad j \in \{-20, \dots, 20\}. \]

\begin{figure}[H]
        \centering
        \includegraphics[width=\textwidth]{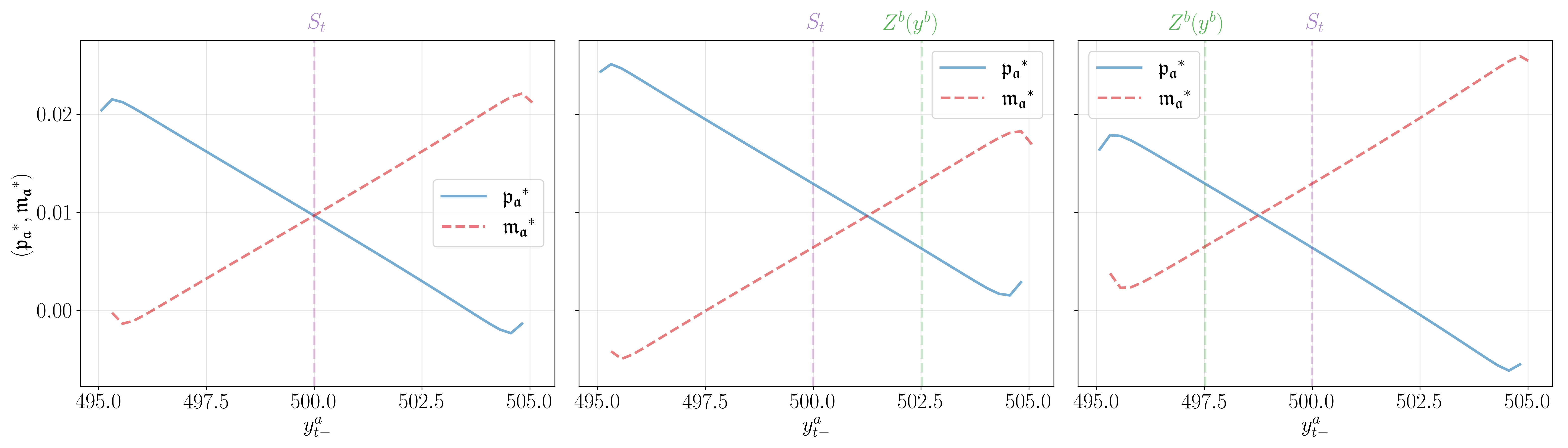}
        \caption{Optimal fees for selling $\feeLTsells^{*}_{a}(t,y_{t-})$ (solid line) and for buying $\feeLTbuys^{*}_{a}(t,y_{t-})$ (dashed line) at time $t=0.5$ as a function of the quantity of asset $Y$ in the pool.}
        \label{fig: Optimal Fees}
\end{figure}

Finally, the time horizon is $T=1$, the oracle price $S_t$ to be $S_0 = 100$ for $t \in [0,T]$ and $y^{a,0} = y^{b,0} = 500$. Figure \ref{fig: Optimal Fees} shows the optimal fees for buying (dotted red line) and selling (continuous blue line) at time $t=0.5$ for player $a$.
This choice of parameters is consistent with that in \cite{baggiani2025optimal}, allowing a direct comparison with the monopoly benchmark. In the latter, all liquidity is concentrated in a single pool, whereas here we analyze a duopoly in which the same aggregate liquidity is allocated across two pools, holding constant the market's aggregate liquidity demand (and thus total traded volume) over the horizon.
The purple line indicates the level of $y^{a}$ for which the instantaneous exchange rate is equal to $S_{t}$ while the green line indicates the instantaneous price of player $b$. In the left plot the instantaneous exchange rate of the opponent $Z^{b}(y^{b})$ is the same as the oracle price, i.e., $Z^{b}(y^{b}) = S_{t} = 100$. This situation corresponds to the case $y^{b} = 500$. In the center plot the opponent exchange rate is $Z^{b}(y^{b}) = 99.1$ corresponding to the quantity $y^{b} = 502$. Finally, in the right plot the opponent exchange rate is $Z^{b}(y^{b}) = 101.1$, corresponding to the quantity $y^{b} = 497$. As in the monopoly case, the fees split into two regimes: one that penalizes arbitrageurs and another that amplifies noise-trader activity (thereby increasing volatility). The difference is that the crossing of the $\feeLTbuys^{a,*}$ and $\feeLTsells^{a,*}$ takes place at a weighted average of the oracle price and the opponent’s exchange rate (as one can see in the second and third panels).

\begin{figure}[H]
        \centering
        \includegraphics[width=\textwidth]{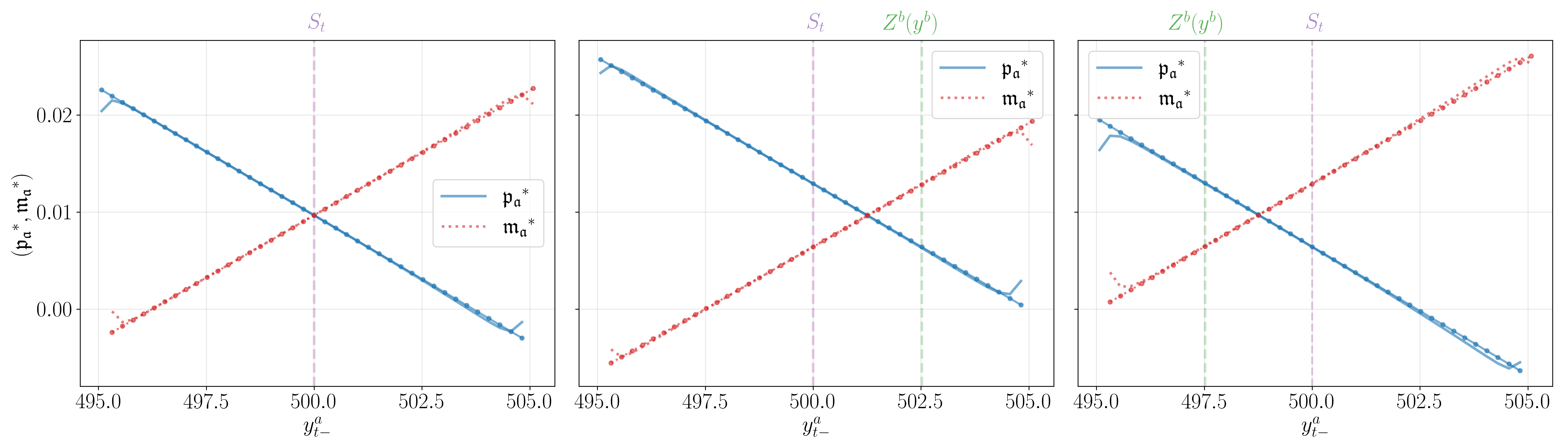}
        \caption{Linear approximation of the fees for selling $\feeLTsells^{*}_{a}(t,y_{t-})$ (solid line) and for buying $\feeLTbuys^{*}_{a}(t,y_{t-})$ (dashed line) at time $t=0.5$ as a function of the quantity of asset $Y$ in the pool.}
        \label{fig: Linear Fees}
\end{figure}

Similarly to the monopoly case, the fees exhibit an approximately linear behaviour in the player's inventory, especially in a neighbourhood of the weighted average of the oracle price and the opponent's exchange rate. Figure \ref{fig: Linear Fees} shows the linear approximations of the fees in the reference agent's inventory. Below, we compare the performance of employing these linear approximations to the optimal fee structure. 

\begin{figure}[H]
    \centering
    \begin{subfigure}[t]{0.48\textwidth}
        \centering
        \includegraphics[width=\textwidth]{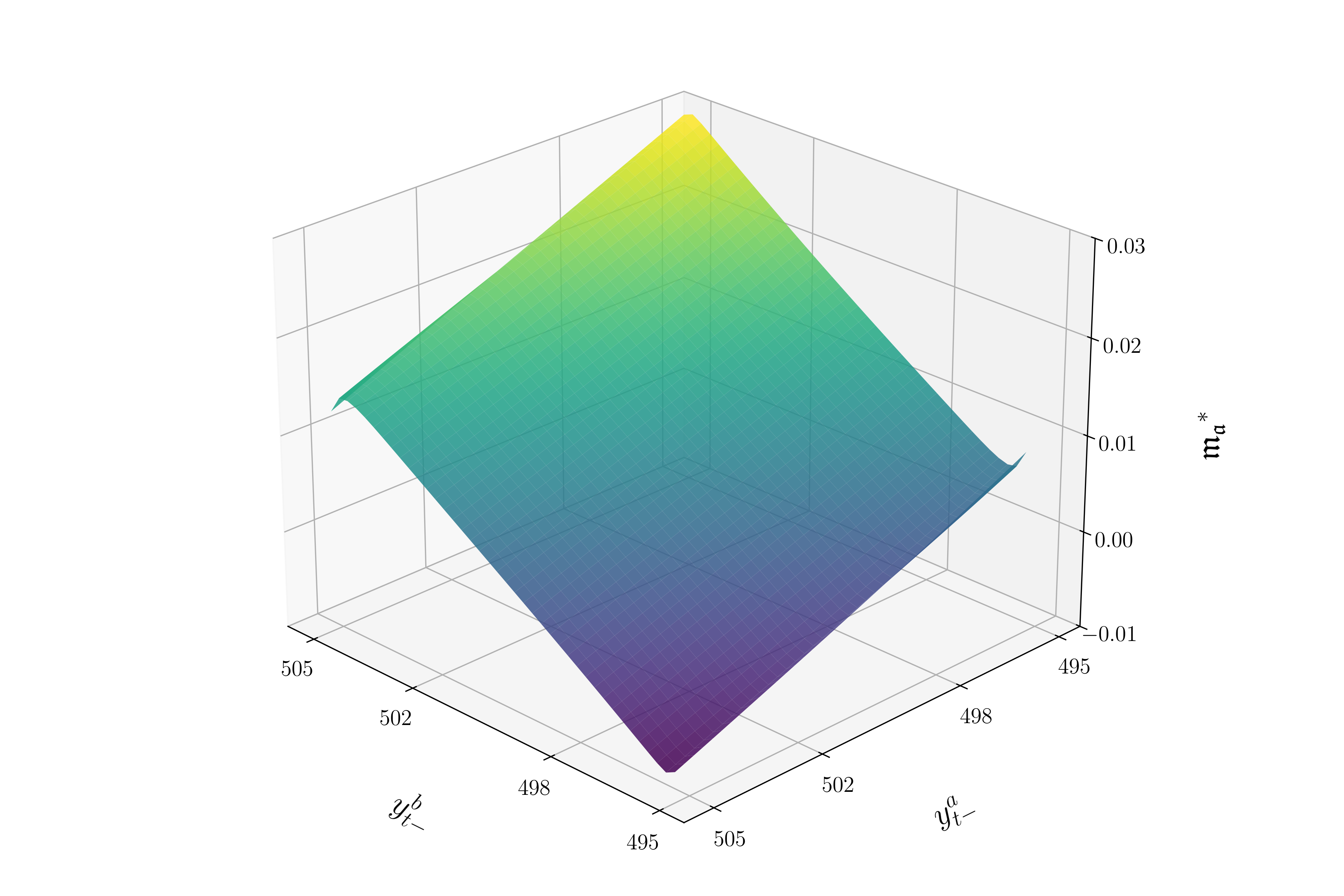}
        \caption{Optimal fees for buying.}
        \label{fig: Optimal fees buy 3d}
    \end{subfigure}
    \hspace*{0.01\textwidth}
    \begin{subfigure}[t]{0.48\textwidth}
        \centering
        \includegraphics[width=\textwidth]{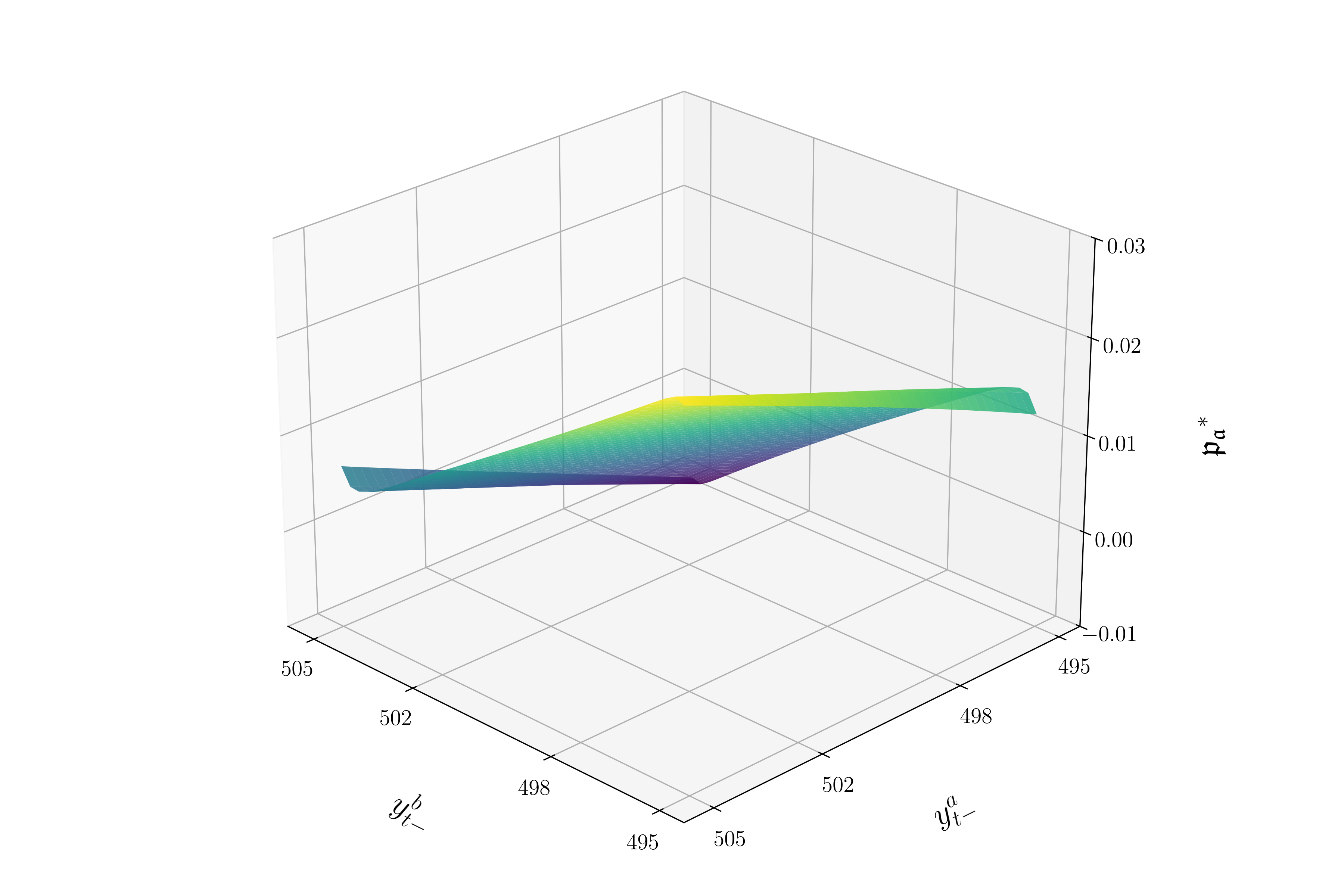}
        \caption{Optimal fees for selling.}
        \label{fig: Optimal fees sell 3d}
    \end{subfigure}
    \caption{Plots of the optimal fees for buying and selling as a function of the agent’s inventory (x-axis) and the opponent’s inventory (y-axis).}
        \label{fig: Optimal 3d Fees}
\end{figure}

Below we show two surfaces of the buy (panel \ref{fig: Optimal fees buy 3d}) and sell (panel \ref{fig: Optimal fees sell 3d}) fees as functions of the agent’s inventory (x-axis) and the opponent’s inventory (y-axis). Here $t=0.5$, $S_{t} = 100$ and we assume that $y^{b} = 500$ (so that $Z^{b}(y^{b}) = 100$)

From these plots, we observe that the fees exhibit an approximately linear dependence on both arguments. The accuracy of this approximation is further supported by the numerical results reported in Tables~\ref{tab:simulations_k2} and \ref{tab:simulations_k1}.

\begin{figure}[H]
        \centering
        \includegraphics[width=\textwidth]{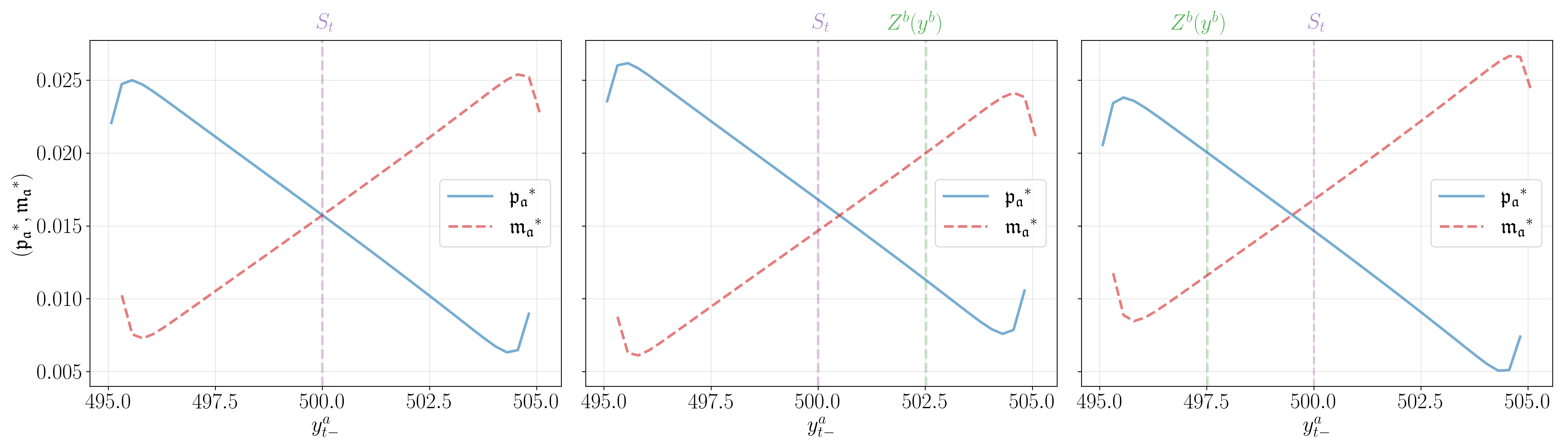}
        \caption{Optimal fees for selling $\feeLTsells^{*}_{a}(t,y_{t-})$ (solid line) and for buying $\feeLTbuys^{*}_{a}(t,y_{t-})$ (dashed line) at time $t=0.5$ as a function of the quantity of asset $Y$ in the pool for $k^{a,b} =0.1$.}
        \label{fig: optimal fees kab = 0.1}
\end{figure}

\begin{figure}[H]
        \centering
        \includegraphics[width=\textwidth]{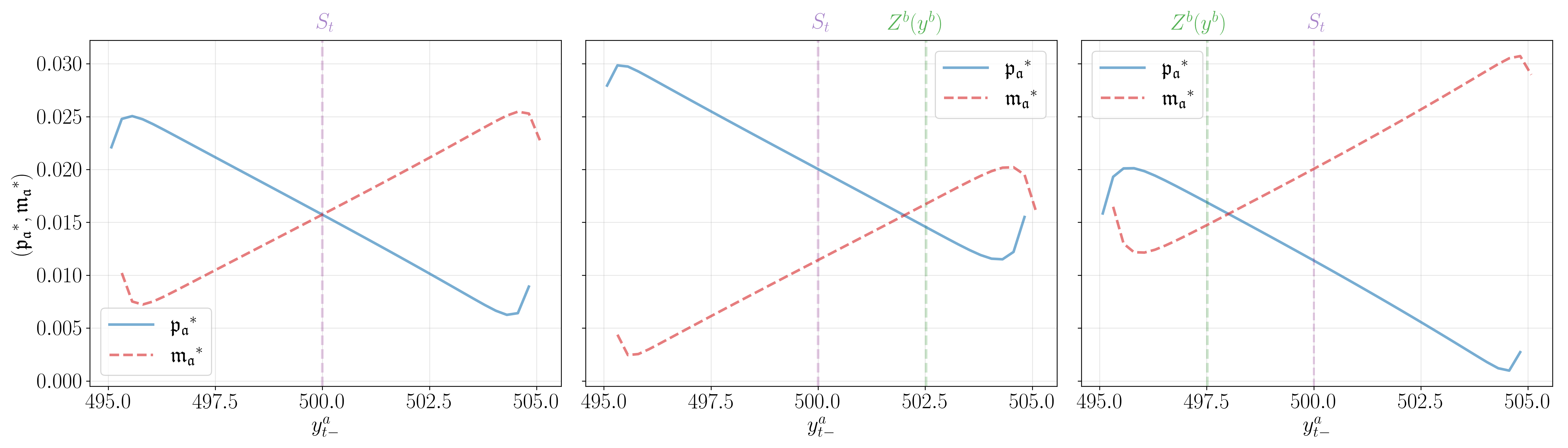}
        \caption{Optimal fees for selling $\feeLTsells^{*}_{a}(t,y_{t-})$ (solid line) and for buying $\feeLTbuys^{*}_{a}(t,y_{t-})$ (dashed line) at time $t=0.5$ as a function of the quantity of asset $Y$ in the pool for $k^{a,0} =0.1$.}
        \label{fig: optimal fees ka0 = 0.1}
\end{figure}

Next, we examine how the optimal fees respond to changes in the sensitivity parameters $k^{a,0}$ and $k^{a,b}$ .
Figure \ref{fig: optimal fees kab = 0.1} shows the buy and sell fees when $k^{a,b}=0.1$.
Two patterns emerge. First, the fee schedules approach the monopoly benchmark: the switching threshold between regimes moves closer to the oracle price. Second, the overall fee levels rise. This follows from the fact that the optimal fees scale inversely with the aggregate sensitivity $k^{a} = k^{a,0}+k^{a,b}$, that is,
\[
\feeLTsells^{a,*},\, \feeLTbuys^{a,*}\ \propto\ \frac{1}{k^{a,0}+k^{a,b}}.
\]
Thus, a reduction in either sensitivity parameter, all else being equal, produces higher fees. This can be seen in Figures \ref{fig: optimal fees kab = 0.1} and \ref{fig: optimal fees ka0 = 0.1}, where we decrease $k^{a,b}$ and $k^{a,0}$ to $0.1$, respectively.

Finally, we examine how the fees respond to changes in the baseline order flow. 
The next plots report the buy and sell fees when $\lambda^{+}=\lambda^{-}=100$.
We observe smoother fee profiles and larger extreme values. 
This follows from the fact that the venue expects more orders so it can afford to charge higher fees.

\begin{figure}[H]
        \centering
        \includegraphics[width=\textwidth]{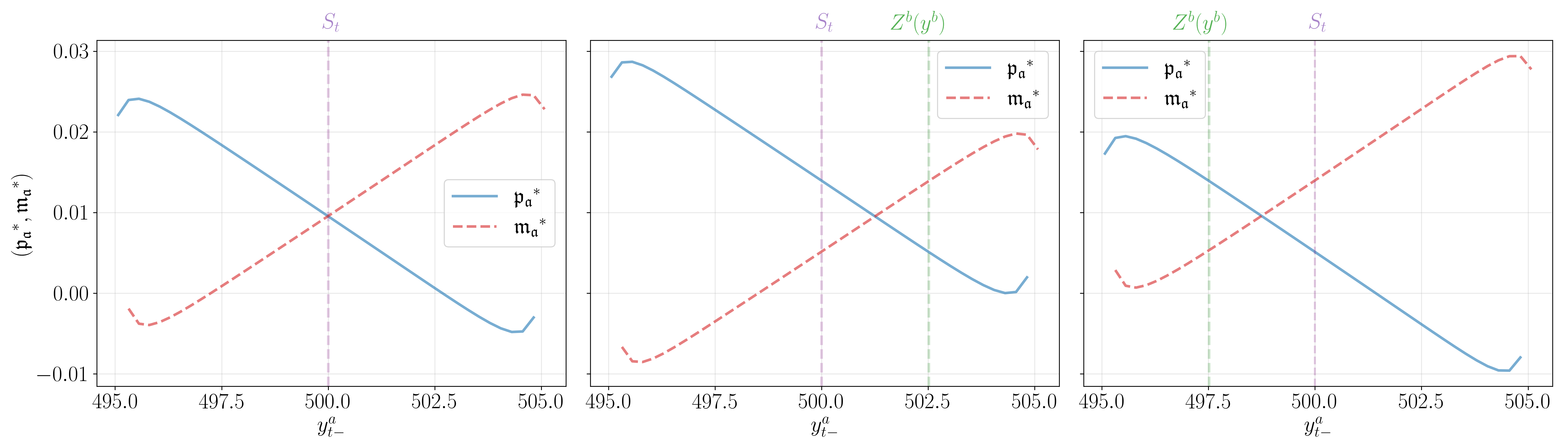}
        \caption{Optimal fees for selling $\feeLTsells^{*}_{a}(t,y_{t-})$ (solid line) and for buying $\feeLTbuys^{*}_{a}(t,y_{t-})$ (dashed line) at time $t=0.5$ as a function of the quantity of asset $Y$ in the pool for $\lambda^{+,a} = \lambda^{-,a} = 100$.}
        \label{fig: optimal fees sensitivity lambda}
\end{figure}

In the following plots we show how the fees behave through time as a function of $y^{a}$, keeping $S$ and $y^{b}$ fixed. The following plots show the behaviour of the fees as a function of time ($x$-axis) and quantity in the pool (colorbar) for different opponent exchange rates. As expected, the fees increase over time.

\begin{figure}[H]
    \centering
    \begin{subfigure}[b]{0.3\textwidth}
        \centering
        \includegraphics[width=\textwidth]{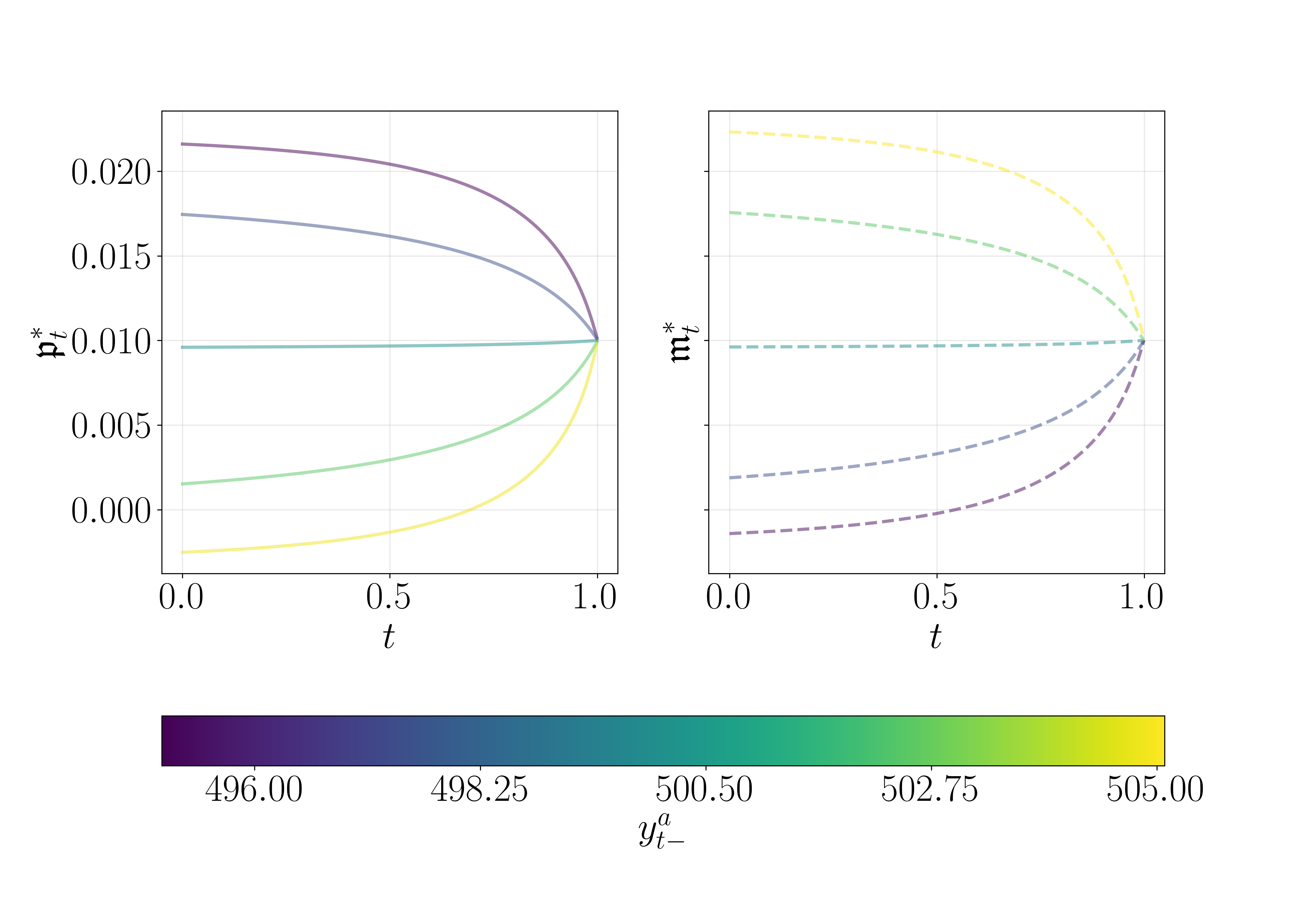}
        \caption{Optimal fees for selling $\feeLTsells^{*}_{t}(y_{t-}^{a},y_{t-}^{b})$ (solid line) and for buying $\feeLTbuys^{*}_{t}(y_{t-}^{a},y_{t-}^{b})$ (dashed line) at time $t=0.5$ as a function of the quantity of asset $Y$ (colorbar) for $y_{t-}^{b} = 500$ and $S = 100$.}
        \label{fig: optimal fees through time}
    \end{subfigure}
    \hfill
    \begin{subfigure}[b]{0.3\textwidth}
        \centering
        \includegraphics[width=\textwidth]{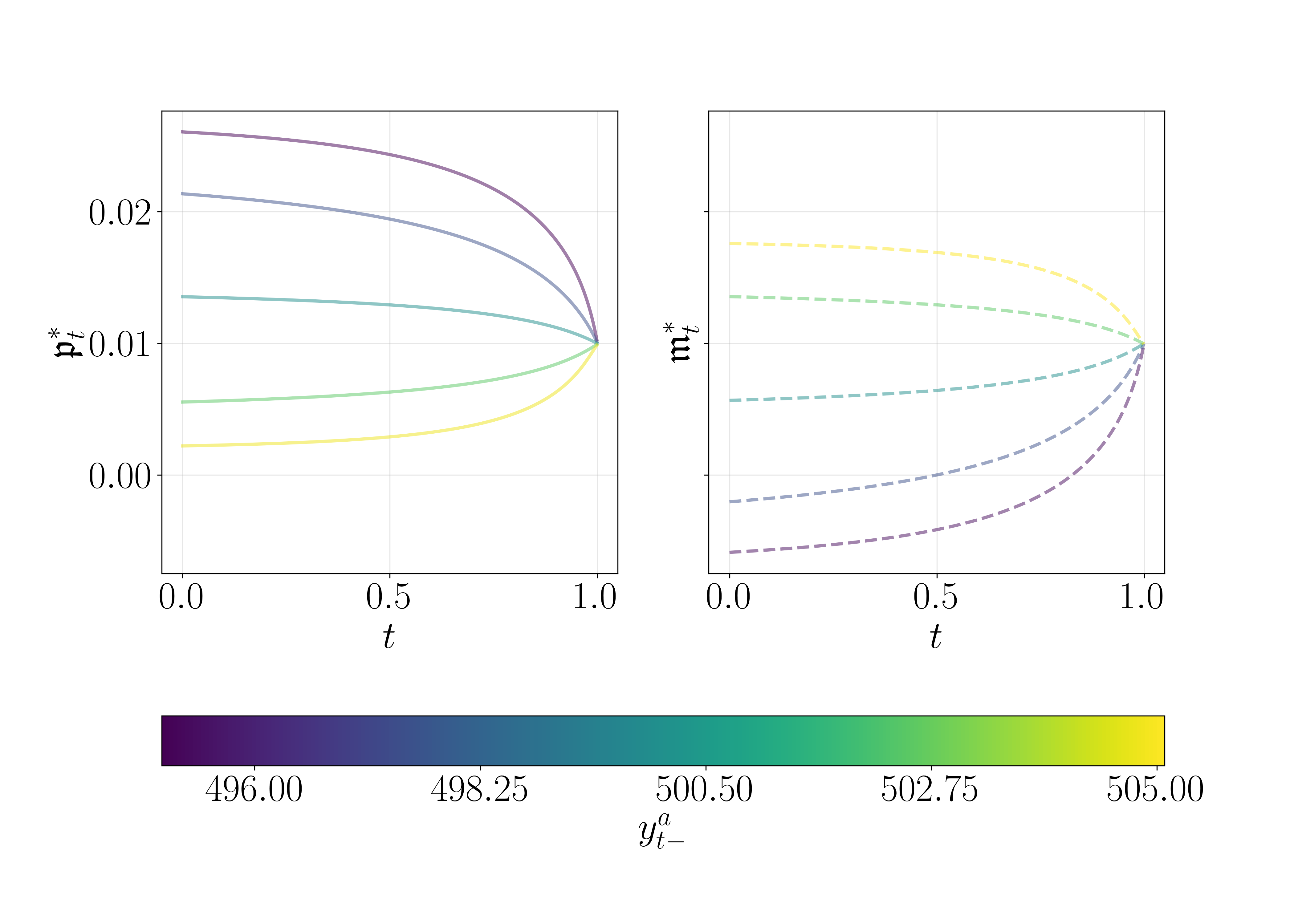}
        \caption{Optimal fees for selling $\feeLTsells^{*}_{t}(y_{t-}^{a},y_{t-}^{b})$ (solid line) and for buying $\feeLTbuys^{*}_{t}(y_{t-}^{a},y_{t-}^{b})$ (dashed line) at time $t=0.5$ as a function of the quantity of asset $Y$ (colorbar) for $y_{t-}^{b} = 502$ and $S = 100$.}
        \label{fig: optimal fees through time imb1}
    \end{subfigure}
    \hfill
    \begin{subfigure}[b]{0.3\textwidth}
        \centering
        \includegraphics[width=\textwidth]{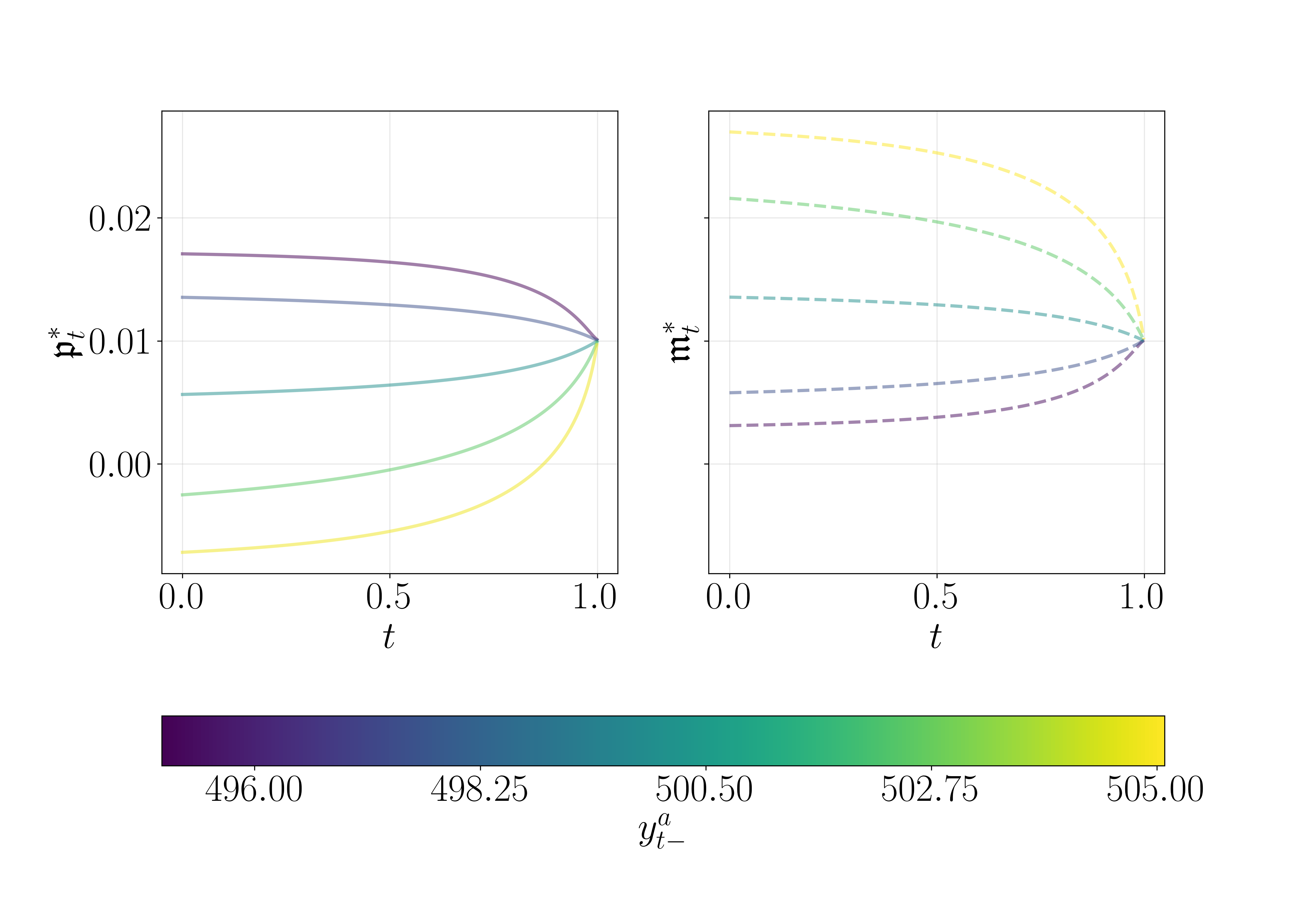}
        \caption{Optimal fees for selling $\feeLTsells^{*}_{t}(y_{t-}^{a},y_{t-}^{b})$ (solid line) and for buying $\feeLTbuys^{*}_{t}(y_{t-}^{a},y_{t-}^{b})$ (dashed line) at time $t=0.5$ as a function of the quantity of asset $Y$ (colorbar) for $y_{t-}^{b} = 497$ and $S = 100$.}
        \label{fig: optimal fees through time imb2}
    \end{subfigure}
\end{figure}

These results resemble the behaviour of the optimal quotes of the market making problem (see Chapter 10 of \cite{cartea2015algorithmic}) and confirm the findings of \cite{baggiani2025optimal}.

Finally, we plot the behaviour of the fees at a fixed time as a function of the centralised price $S_{t}$. We observe that the optimal fees are monotone in the oracle price $S_t$ as the buy fee $\mathfrak{m}_t^\ast$ increases with $S_t$, while the sell fee $\mathfrak{p}_t^\ast$ decreases.
Intuitively, when $S_t$ rises, buy-side arbitrage/flow into the pool becomes more attractive, so the venue raises
$\mathfrak{m}_t^\ast$ to make buying more expensive and reduce that pressure; conversely it lowers $\mathfrak{p}_t^\ast$
to keep the pool competitive for incoming sell orders. The opponent's inventory shifts the level of the curves because it changes the opponent's quote, therefore changing the weighted average of the centralised price and the opponent's price.
The curves can intersect at values of $S_t$ where two different inventory states imply the same effective mispricing.

\begin{figure}[H]
    \centering
    \begin{subfigure}[b]{0.32\textwidth}
        \centering
        \includegraphics[width=\textwidth]{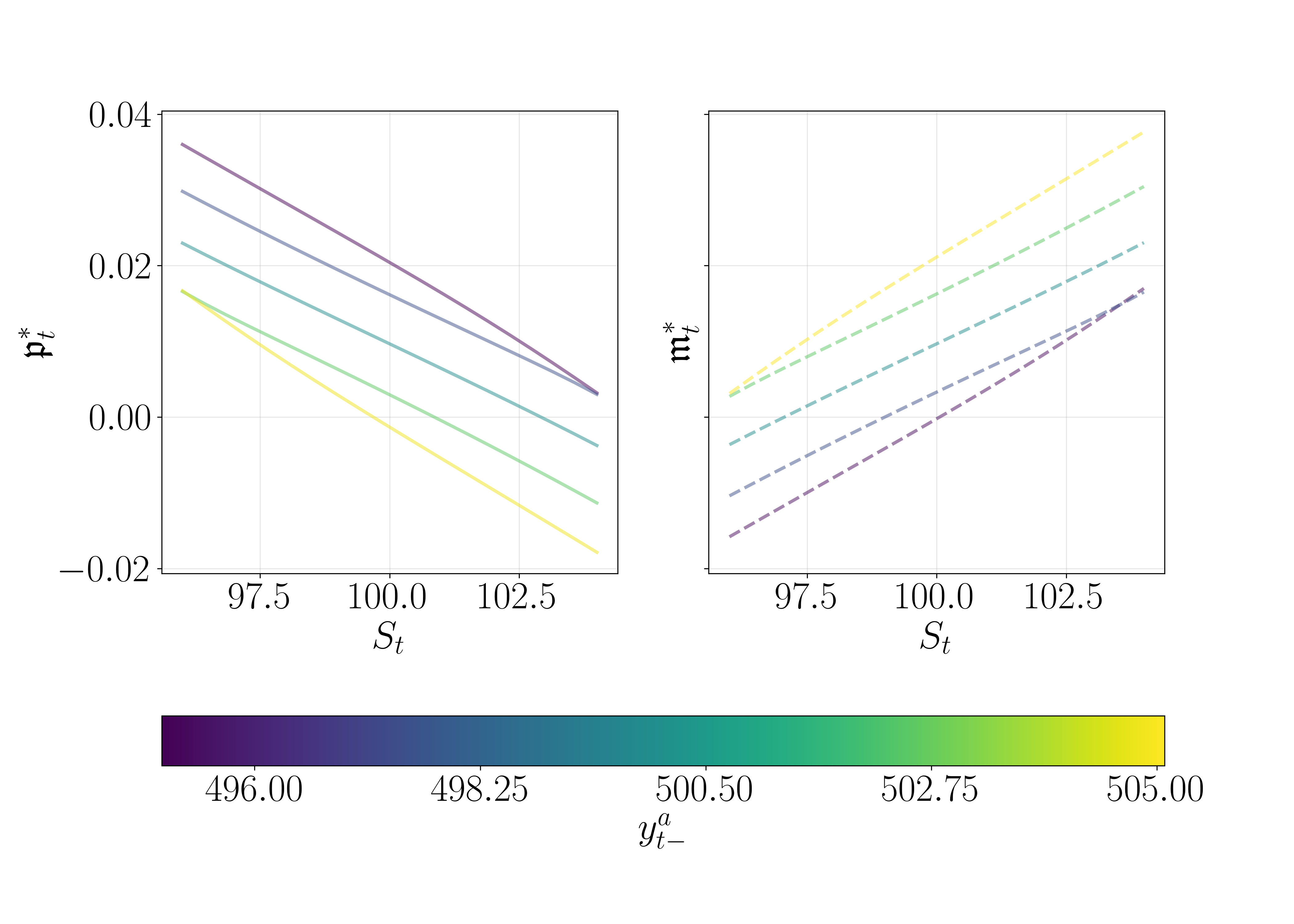}
        \caption{Optimal fees for selling $\feeLTsells^{*}_{t}(y_{t-}^{a},y_{t-}^{b})$ (solid line) and for buying $\feeLTbuys^{*}_{t}(y_{t-}^{a},y_{t-}^{b})$ (dashed line) at time $t=0.5$ as functions of the oracle price $S_t$ and $y_{t-}^{a}$ (colorbar) for $y_{t-}^{b}=500$.}
        \label{fig:fees-St-yb500}
    \end{subfigure}
    \hfill
    \begin{subfigure}[b]{0.32\textwidth}
        \centering
        \includegraphics[width=\textwidth]{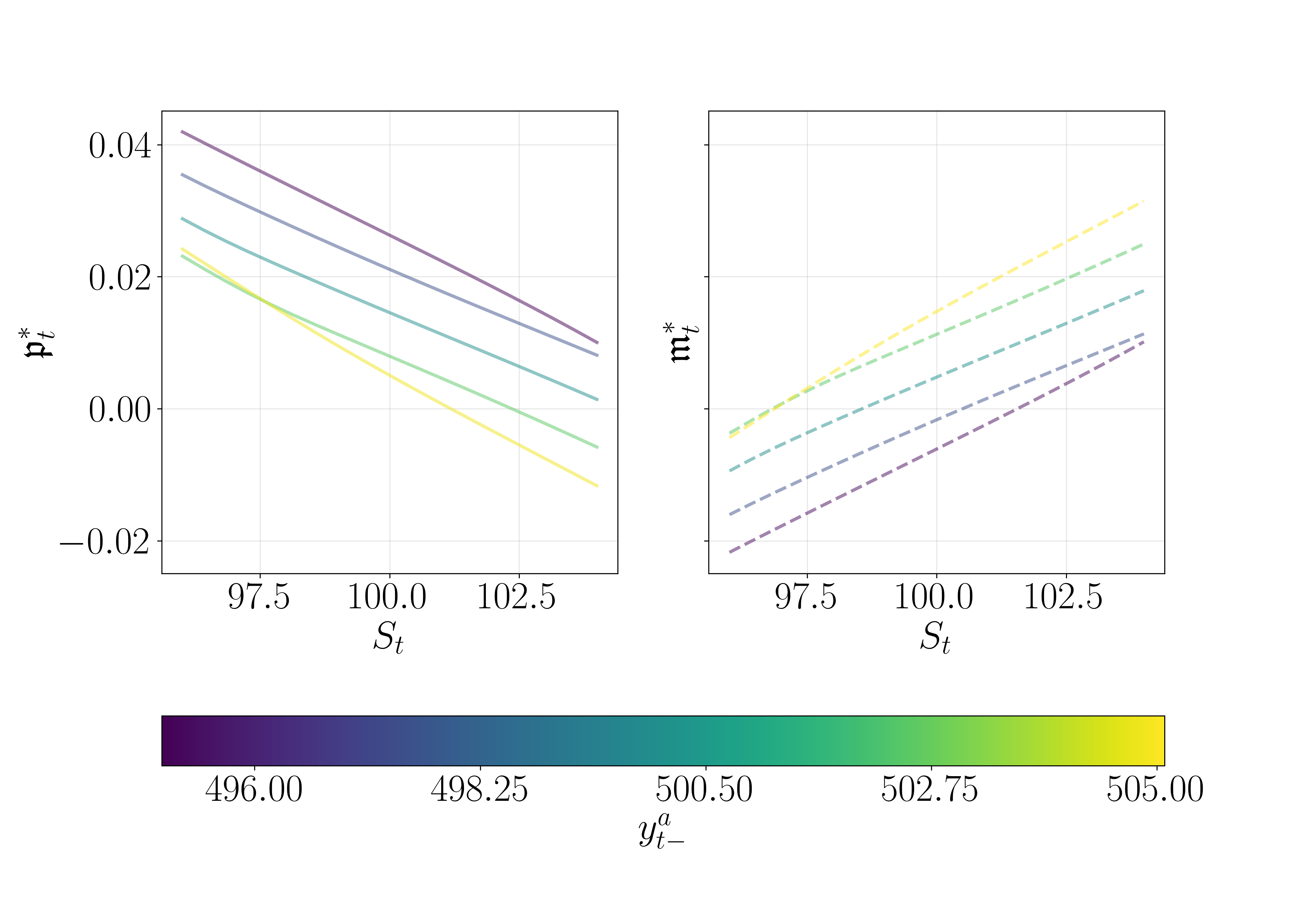}
        \caption{Optimal fees for selling $\feeLTsells^{*}_{t}(y_{t-}^{a},y_{t-}^{b})$ (solid line) and for buying $\feeLTbuys^{*}_{t}(y_{t-}^{a},y_{t-}^{b})$ (dashed line) at time $t=0.5$ as functions of the oracle price $S_t$ and $y_{t-}^{a}$ (colorbar) for $y_{t-}^{b}=503$.}
        \label{fig:fees-St-yb503}
    \end{subfigure}
    \hfill
    \begin{subfigure}[b]{0.32\textwidth}
        \centering
        \includegraphics[width=\textwidth]{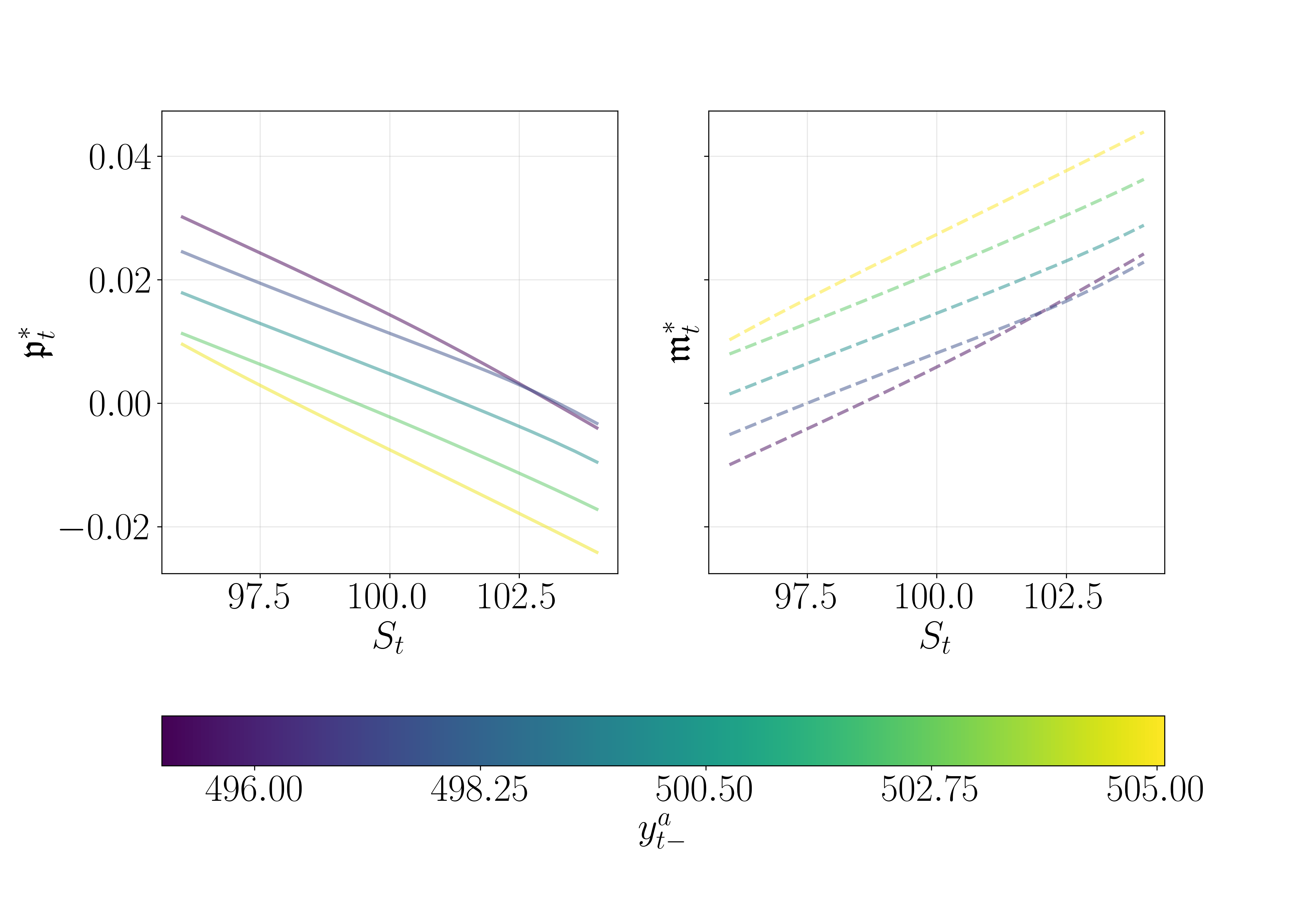}
        \caption{Optimal fees for selling $\feeLTsells^{*}_{t}(y_{t-}^{a},y_{t-}^{b})$ (solid line) and for buying $\feeLTbuys^{*}_{t}(y_{t-}^{a},y_{t-}^{b})$ (dashed line) at time $t=0.5$ as functions of the oracle price $S_t$ and $y_{t-}^{a}$ (colorbar) for $y_{t-}^{b}=497$.}
        \label{fig:fees-St-yb497}
    \end{subfigure}
    \label{fig:fees-vs-St-competition}
\end{figure}

\section{Simulations} \label{section: two player model simulations}

We now perform numerical simulations in a two-player setting and compare the results with the monopoly case. To ensure a \emph{fair} comparison with the monopoly case the total depth is split between the two pools, and for simplicity we assume complete symmetry between the two pools.
The grid for the exchange rate $Z$ is the same in both the monopoly and duopoly cases. In this case we set $k^{i,0} = k^{i,j} = k$, where $k$ is the exponential decay parameter in the monopoly case. Moreover, we impose $\lambda^{i,+} = \lambda^{+}$ and $\lambda^{i,-} = \lambda^{-}$.
If we take the grid for the risky asset $y$ as
\[
y^{i,j} := \sqrt{\frac{(p^{i})^{2}}{Z^{i}(y^{i,0}) - 0.1 \, j}},
\qquad i \in \{a,b\}, \quad j \in \{-20, \dots, 20\},
\]
then the resulting trade size is roughly half that of the monopoly case.
In this scenario, each venue receives a similar number of orders as in the monopoly case; however, each order is roughly half the size, so that the total traded volume (i.e. the sum of the volume traded in the first and in the second venue) is comparable to the total volume traded in the monopoly case. Tables~\ref{tab:simulations_k2} and \ref{tab:simulations_k1} show the results of the simulations for the monopoly and duopoly cases with the parameters above. The number of simulations is $100,000$ and $S_t = 100$ for simplicity. We compare the Nash equilibrium against the strategies:

\begin{itemize}
    \item[$(i)$] the \emph{linear} strategy where both $\feeLTsells^{*}_{i}$ and $\feeLTbuys^{*}_{i}$ are linearized around a neighborhood of $(y^{0,a}, y^{0,b})$ and
    \item[$(ii)$] the \emph{constant} strategy where both players set constant fees every time $t$ and every quantity $y$. The constants $c^{i}$ are chosen as the average of the optimal two fees of player $i$ at $t=0.5$ for $y^{i} = y^{i,0}$ and $y^{j} = y^{j,0}$, i.e., \[ c^{i} := \frac{\feeLTsells^{*}_{i}(0.5,y^{a,0},y^{b,0}) + \feeLTbuys^{*}_{i}(0.5,y^{a,0},y^{b,0})}{2}. \]
\end{itemize}
We see that the Nash equilibrium outperforms the constant case by between $50$ and $150$ bps. We see that the linear approximation of the fees provides an almost perfect approximation, indistinguishable after integer rounding. This is because trading takes place in a neighborhood of $(y^{0,a}, y^{0,b})$, as arbitrage aligns the pool prices with the oracle price $S_{t}$.
Finally, we see that the total values are similar between the monopoly and the two-player case.

\begin{table}[H]
\centering
\begin{tabular}{@{}cccccccccc@{}}
\toprule
& & \multicolumn{4}{c}{$\lambda = 100$} & \multicolumn{4}{c}{$\lambda = 150$} \\
\cmidrule(lr){3-6} \cmidrule(lr){7-10}
Player & Type & Fees & Sell & Buy & Vol & Fees & Sell & Buy & Vol \\
\midrule
\multirow{3}{*}{A} 
 & Optimal  & 18.22 & 36.79 & 36.78 & 1839 & 27.13 & 54.77 & 54.75 & 2737 \\
 & Linear   & 18.22 & 36.79 & 36.78 & 1839 & 27.13 & 54.77 & 54.75 & 2737 \\
 & Constant & 17.99 & 37.45 & 37.46 & 1872 & 26.72 & 55.80 & 55.81 & 2790 \\
\cmidrule(lr){1-10}
\multirow{3}{*}{B} 
 & Optimal  & 18.23 & 36.79 & 36.79 & 1839 & 27.13 & 54.77 & 54.73 & 2737 \\
 & Linear   & 18.23 & 36.79 & 36.79 & 1839 & 27.13 & 54.77 & 54.73 & 2737 \\
 & Constant & 17.99 & 37.44 & 37.45 & 1872 & 26.71 & 55.78 & 55.78 & 2789 \\
\cmidrule(lr){1-10}
\multirow{3}{*}{Total} 
 & Optimal  & 36.45 & 73.58 & 73.57 & 3678 & 54.26 & 109.54 & 109.48 & 5474 \\
 & Linear   & 36.45 & 73.58 & 73.57 & 3678 & 54.26 & 109.54 & 109.48 & 5474 \\
 & Constant & 35.98 & 74.89 & 74.91 & 3744 & 53.43 & 111.58 & 111.59 & 5579 \\
\cmidrule(lr){1-10}
\multirow{3}{*}{Monopoly} 
 & Optimal  & 35.55 & 35.89 & 35.91 & 3590 & 52.97 & 53.45 & 53.47 & 5345 \\
 & Linear   & 35.55 & 35.88 & 35.91 & 3589 & 52.97 & 53.44 & 53.48 & 5345 \\
 & Constant & 35.16 & 35.15 & 35.15 & 3653 & 52.26 & 52.26 & 52.26 & 5445 \\
\bottomrule
\end{tabular}
\caption{Average fees collected (Fees), number of sell trades (Sell), number of buy trades (Buy), and volume traded (Vol) for players $A$ and $B$ under different fee types when $k=2$, with $\lambda\in\{100,150\}$.}
\label{tab:simulations_k2}
\end{table}

As discussed in Remark~\ref{rem: optimal fees connection with k}, optimal fee revenues scale approximately like $1/k$; hence, holding everything else fixed, increasing $k$ from $1$ to $2$ leads to roughly half the fee revenue. Moreover, when $\lambda$ increases there is more order flow, so it is natural that optimal fees (and revenues) increase with $\lambda$, we show this in Table \ref{tab:simulations_k1}.

\begin{table}[H]
\centering
\begin{tabular}{@{}cccccccccc@{}}
\toprule
& & \multicolumn{4}{c}{$\lambda = 100$} & \multicolumn{4}{c}{$\lambda = 150$} \\
\cmidrule(lr){3-6} \cmidrule(lr){7-10}
Player & Type & Fees & Sell & Buy & Vol & Fees & Sell & Buy & Vol \\
\midrule
\multirow{3}{*}{A} 
 & Optimal  & 36.18 & 36.37 & 36.37 & 1818 & 53.85 & 54.12 & 54.13 & 2706 \\
 & Linear   & 36.18 & 36.37 & 36.37 & 1818 & 53.85 & 54.12 & 54.13 & 2706 \\
 & Constant & 36.04 & 36.62 & 36.64 & 1831 & 53.56 & 54.54 & 54.58 & 2728 \\
\cmidrule(lr){1-10}
\multirow{3}{*}{B} 
 & Optimal  & 36.19 & 36.37 & 36.38 & 1818 & 53.84 & 54.11 & 54.11 & 2705 \\
 & Linear   & 36.19 & 36.37 & 36.38 & 1818 & 53.84 & 54.11 & 54.11 & 2705 \\
 & Constant & 36.04 & 36.62 & 36.64 & 1831 & 53.54 & 54.53 & 54.55 & 2727 \\
\cmidrule(lr){1-10}
\multirow{3}{*}{Total} 
 & Optimal  & 72.37 & 72.74 & 72.75 & 3636 & 107.69 & 108.23 & 108.24 & 5411 \\
 & Linear   & 72.37 & 72.74 & 72.75 & 3636 & 107.69 & 108.23 & 108.24 & 5411 \\
 & Constant & 72.08 & 73.24 & 73.28 & 3662 & 107.10 & 109.07 & 109.13 & 5455 \\
\cmidrule(lr){1-10}
\multirow{3}{*}{Monopoly} 
 & Optimal  & 71.46 & 35.92 & 35.94 & 3593 & 106.38 & 53.47 & 53.50 & 5348 \\
 & Linear   & 71.46 & 35.91 & 35.95 & 3593 & 106.38 & 53.46 & 53.50 & 5348 \\
 & Constant & 71.22 & 35.60 & 35.62 & 3618 & 105.90 & 52.94 & 52.96 & 5390 \\
\bottomrule
\end{tabular}
\caption{Average fees collected (Fees), number of sell trades (Sell), number of buy trades (Buy), and volume traded (Vol) for players $A$ and $B$ under different fee types when $k=1$, with $\lambda\in\{100,150\}$.}
\label{tab:simulations_k1}
\end{table}

Next we analyze the benefits of competition for the different parties involved. We first consider the point of view of a liquidity taker. Our goal is to understand whether a liquidity taker is better off trading in a  world with two competing  \emph{venues} (playing the Nash equilibrium of the game), each with $500$ units of asset $Y$, or with only one with a total of $1000$ units of asset $Y$.

To do so, we consider two scenarios: (i) a strategic liquidity taker wishing to execute a trade of a fixed size $D$, and (ii) the liquidity takers that arrive to the venues according to the stochastic intensities of the model. For the first scenario, when there is only one venue, the liquidity taker will send the trade of size $D$ to the venue and pay the corresponding fee. For the second scenario, the trade of size $D$ is split in two trades of size $D/2$ and either he sends both trades to the first venue, both trades to the second venue, or one trade to each venue, depending on what is more advantageous at the time of sending the trades.  

\begin{figure}[H]
    \centering
    \begin{subfigure}[t]{0.48\textwidth}
        \centering
        \includegraphics[width=\textwidth]{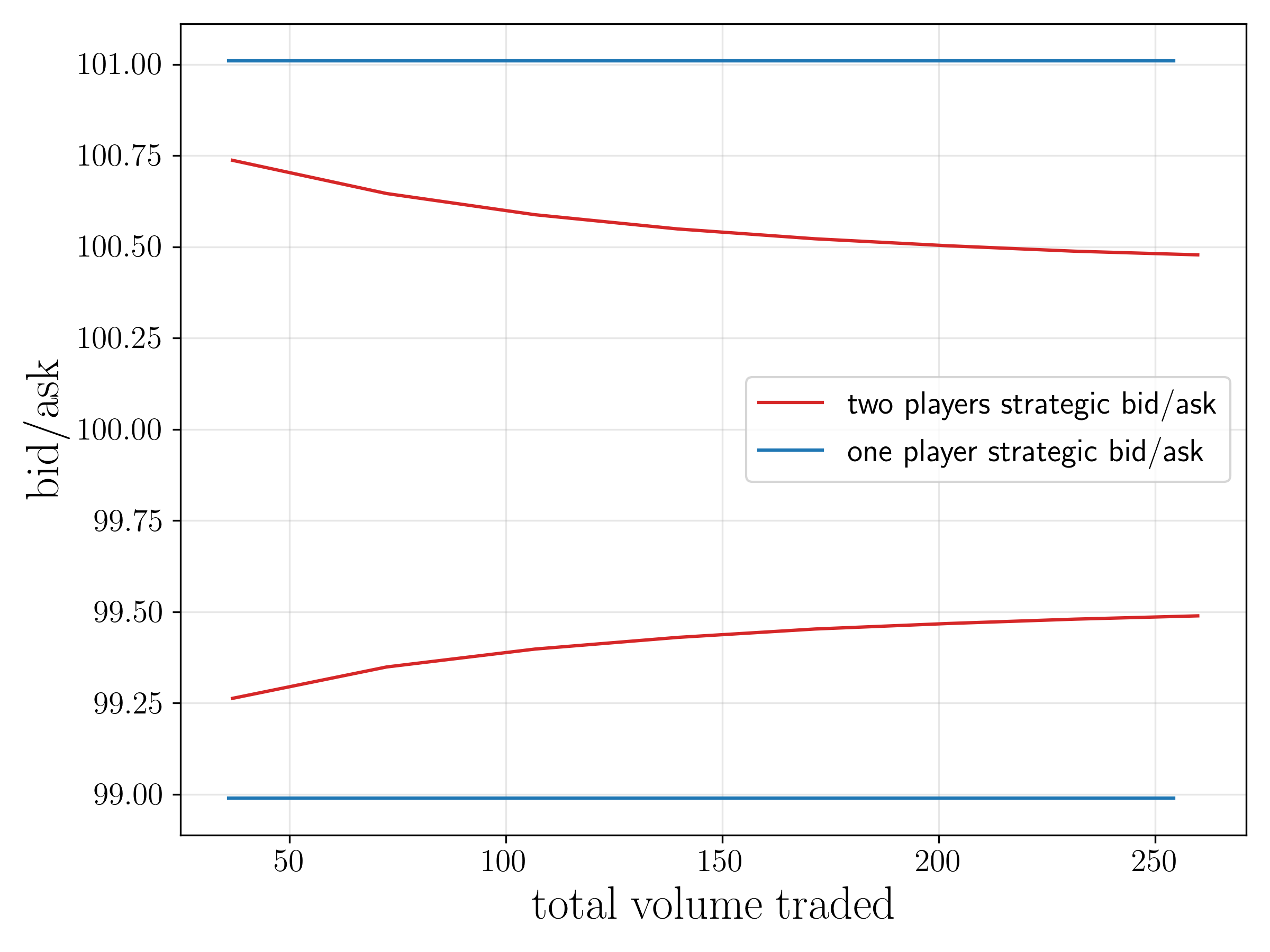}
        \caption{Strategic Bid--ask as a function of total traded volume in the monopoly (blue line) and duopoly (red line) case.}
        \label{fig:bid-ask-2players}
    \end{subfigure}
    \hspace*{0.01\textwidth}
    \begin{subfigure}[t]{0.48\textwidth}
        \centering
        \includegraphics[width=\textwidth]{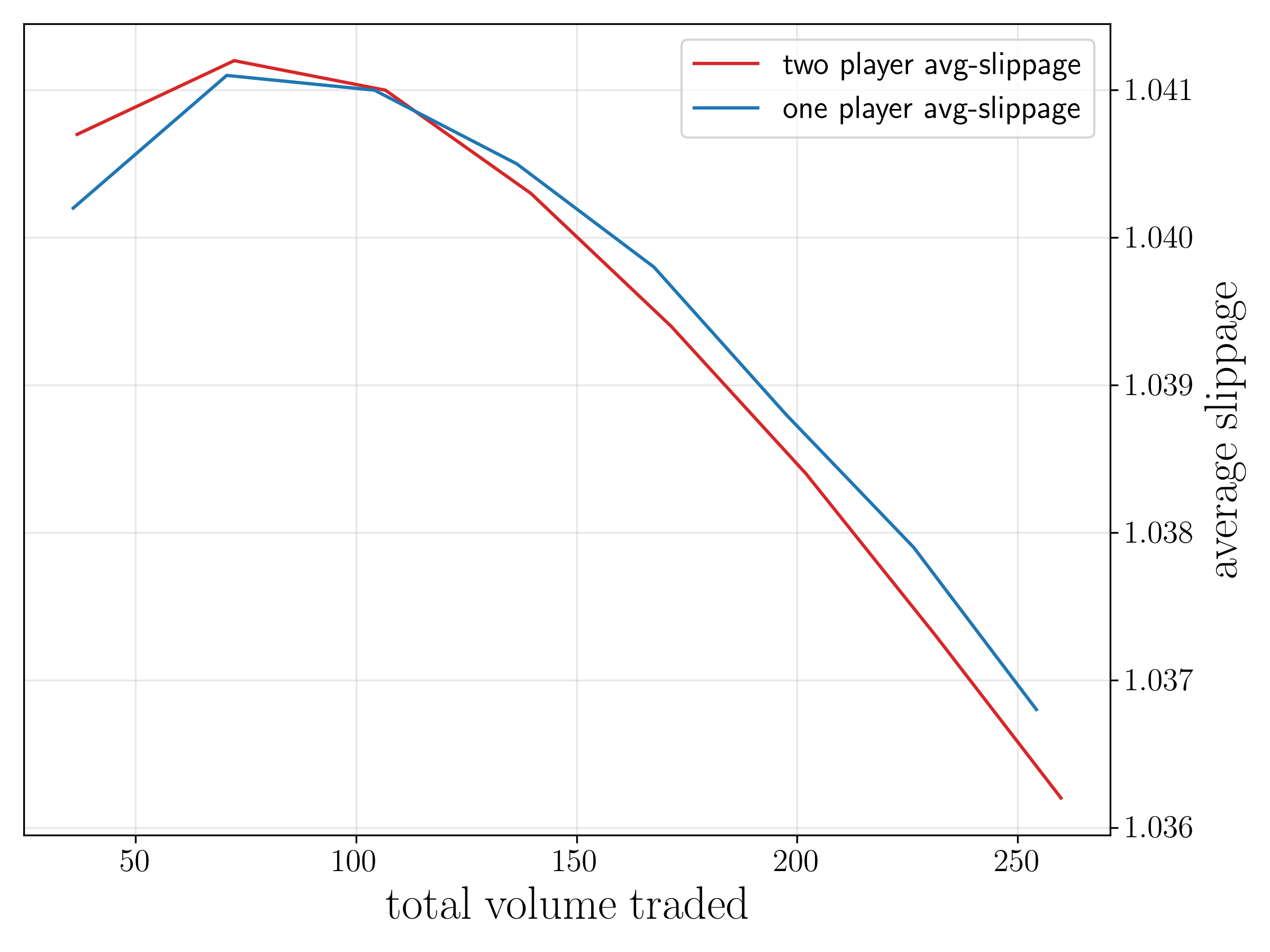}
        \caption{Average slippage as a function of total traded volume in the monopoly (blue line) and duopoly (red line) case.}
        \label{fig:slippage-2players}
    \end{subfigure}
\end{figure}

Second, we compute the \emph{average slippage}, defined as the absolute deviation of the execution quote from the marginal exchange rate divided by the total traded volume. More precisely, we have that\footnote{Equivalently, slippage is the absolute distance of the fee-adjusted bid/ask rate from the oracle price at the trade time.}  
\begin{align*}
    \text{Avg-slippage} := \mathbb{E} \left[ \frac{ \sum_{i \in \{a,b\}} \int_{0}^{T}(Z^{i}(y^{i}_{t}) - \exratesell^{i,\feeLTsells^*}(y^{i}_{t})) \Delta^{i}_{+}(y^{i}) \dd N^{i,+}_{t}  + \int_{0}^{T}(\exratebuy^{i,\feeLTbuys^*}(y^{i}_{t}) - Z^{i}(y^{i}_{t})) \Delta^{i}_{-}(y^{i}) \dd N^{i,-}_{t}}{\sum_{i \in \{a,b\}} \int_{0}^{T} \Delta^{i}_{+}(y^{i}_{t}) \dd N^{i,+}_{t} + \int_{0}^{T} \Delta^{i}_{-}(y^{i}_{t}) \dd N^{i,-}_{t} } \right].
\end{align*}
We plot both quantities as a function of the market demand for liquidity, which we hold fixed and use as the $x$-axis. From Figure~\eqref{fig:bid-ask-2players} we observe two phenomena. First, competition benefits \emph{strategic} liquidity takers who decide to trade at the best available price. Second, in the two-player case, the average \emph{best} spread decreases as traded volume increases. We explain the tightening of the \emph{best} quote under competition through order statistics. The best ask in the duopoly is a minimum across pools, while the best bid is a maximum across pools; increasing the dispersion of cross-sectional quotes mechanically decreases the expected minimum and increases the expected maximum. This is formalized by the following remark.

\begin{remark}
If $X$ and $Y$ two i.i.d. random variables such that $\mathbb{E}[\vert X \vert^2] < \infty$ with mean $\mu$ and variance $\sigma^{2}$ then the quantity $\mathbb{E}[\min(X,Y)]$ (resp. $\mathbb{E}[\max(X,Y)]$) is decreasing (resp. increasing) as a function of the standard deviation $\sigma$. 
\end{remark}

The average slippage in Figure~\eqref{fig:slippage-2players} presents two distinct features. First, we notice that we can decompose the numerator of the average slippage as the convexity charge (which is not affected by the fees),
\[
\sum_{i \in \{a,b\}} \int_{0}^{T}\bigl(Z^{i}(y^{i}_{t}) - \exratesell^{i}(y^{i}_{t})\bigr)\,\Delta^{i}_{+}(y^{i}) \,\dd N^{i,+}_{t}
\;+\;
\int_{0}^{T}\bigl(\exratebuy^{i}(y^{i}_{t}) - Z^{i}(y^{i}_{t})\bigr)\,\Delta^{i}_{-}(y^{i}) \,\dd N^{i,-}_{t},
\]
plus the total cash collected (which is affected by the fees and hence differs between the one- and two-player cases),
\[
\int_0^t \feeLTsells_{u}^{i}\,\exratesell^{i,\feeLTsells_{u}^{i}}\!\bigl(Y_{u}^{i,\fees}\bigr)\,\Delta_{+}^{i}\!\bigl(Y_{t}^{i,\fees}\bigr)\,\dd N^{i,+,\feeLTsells^{i}}_{u}
\;+\;
\int_0^t \feeLTbuys^{i}_{u}\,\exratebuy^{i,\feeLTbuys^{i}_{u}}\!\bigl(Y_{u}^{i,\fees}\bigr)\,\Delta_{-}^{i}\!\bigl(Y_{t}^{i,\fees}\bigr)\,\dd N^{i,+,\feeLTbuys^{i}}_{u}.
\]

We next observe that the slippage is concave with respect to the total volume traded. It is decreasing because higher volume implies higher absolute cash revenue but lower relative cash revenue. This is because more volume traded implies higher volatility, which can be beneficial for the LP up to a certain point. However, a very high-volatility regime implies that the asset reaches the boundaries more often, and the boundaries are where the AMM does not charge any fees; this explains the concavity. In the two-player case, the number of times the asset reaches the inventory is more than double that of the one-player case, so the inventory is closer to the no-trading zone. Hence, in the two-player case the two players collect less money per unit of volume traded.

Finally, we analyze whether the venue and the liquidity providers benefit from competition. We consider a setting in which a single venue (e.g.\ Uniswap) hosts both the monopoly and the two-player configurations and collects $10\%$ of the total fees earned by the liquidity providers in each case. In our setting, the venue is essentially indifferent to competition: because it collects a fixed amount of LP fees and total fees are unchanged between monopoly and duopoly, its revenue is the same in both cases. However, total revenue for a single LP declines under competition, since trading flow (and thus fee generation) is split across the two pools rather than concentrated in a single one. Combining these results with the previous one, we find that greater competition benefits strategic liquidity takers, while noise liquidity takers bear the cost.

\begin{figure}[H]
    \centering
    \begin{subfigure}[t]{0.48\textwidth}
        \centering
        \includegraphics[width=\textwidth]{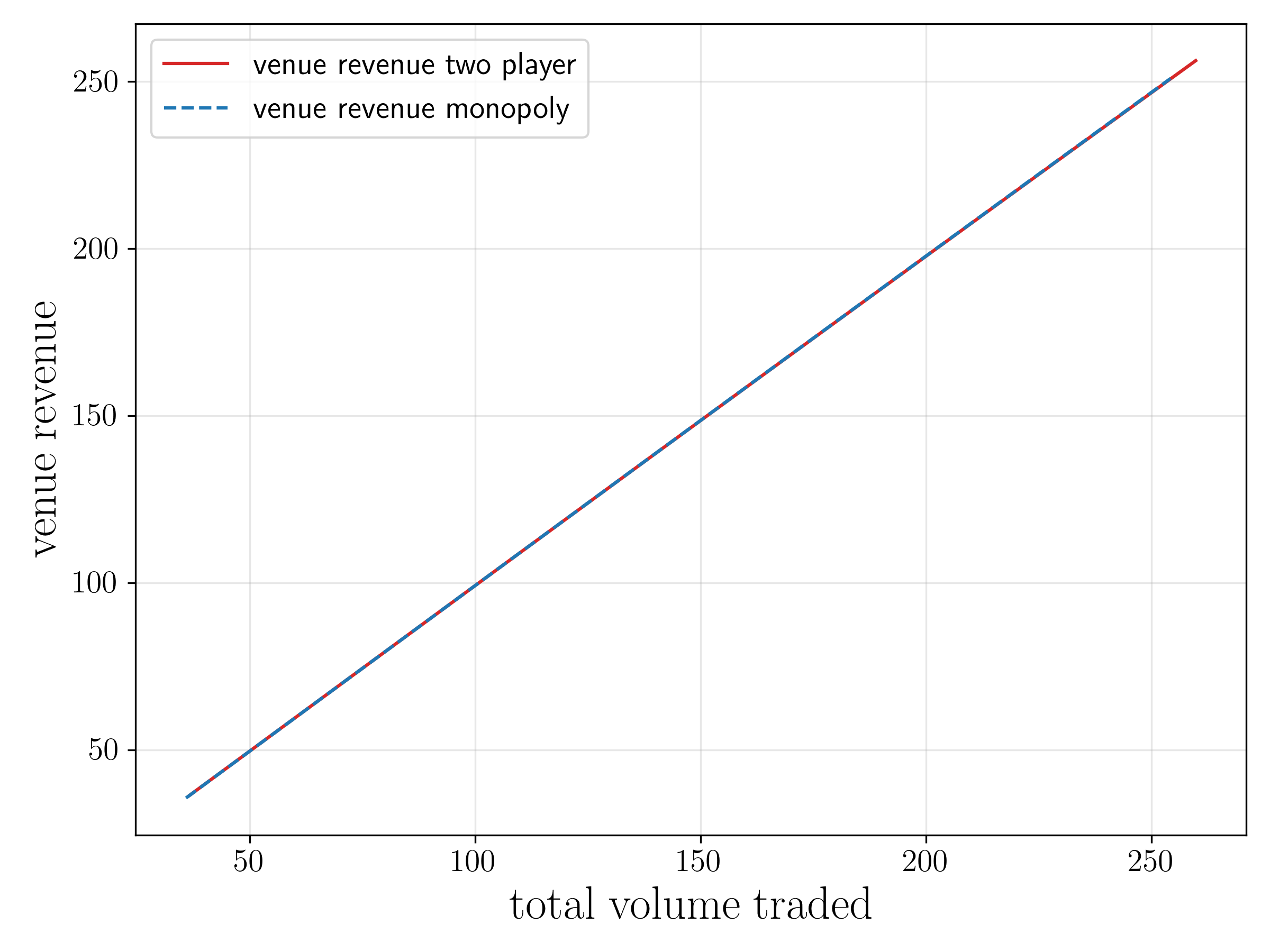}
        \caption{Total revenue of the venue computed as the sum of a fixed percentage of the fee revenue of the players in the monopoly (blue line) and two-player (red line) case.}
        \label{fig:total revenue 2-p}
    \end{subfigure}
    \hspace*{0.01\textwidth}
    \begin{subfigure}[t]{0.48\textwidth}
        \centering
        \includegraphics[width=\textwidth]{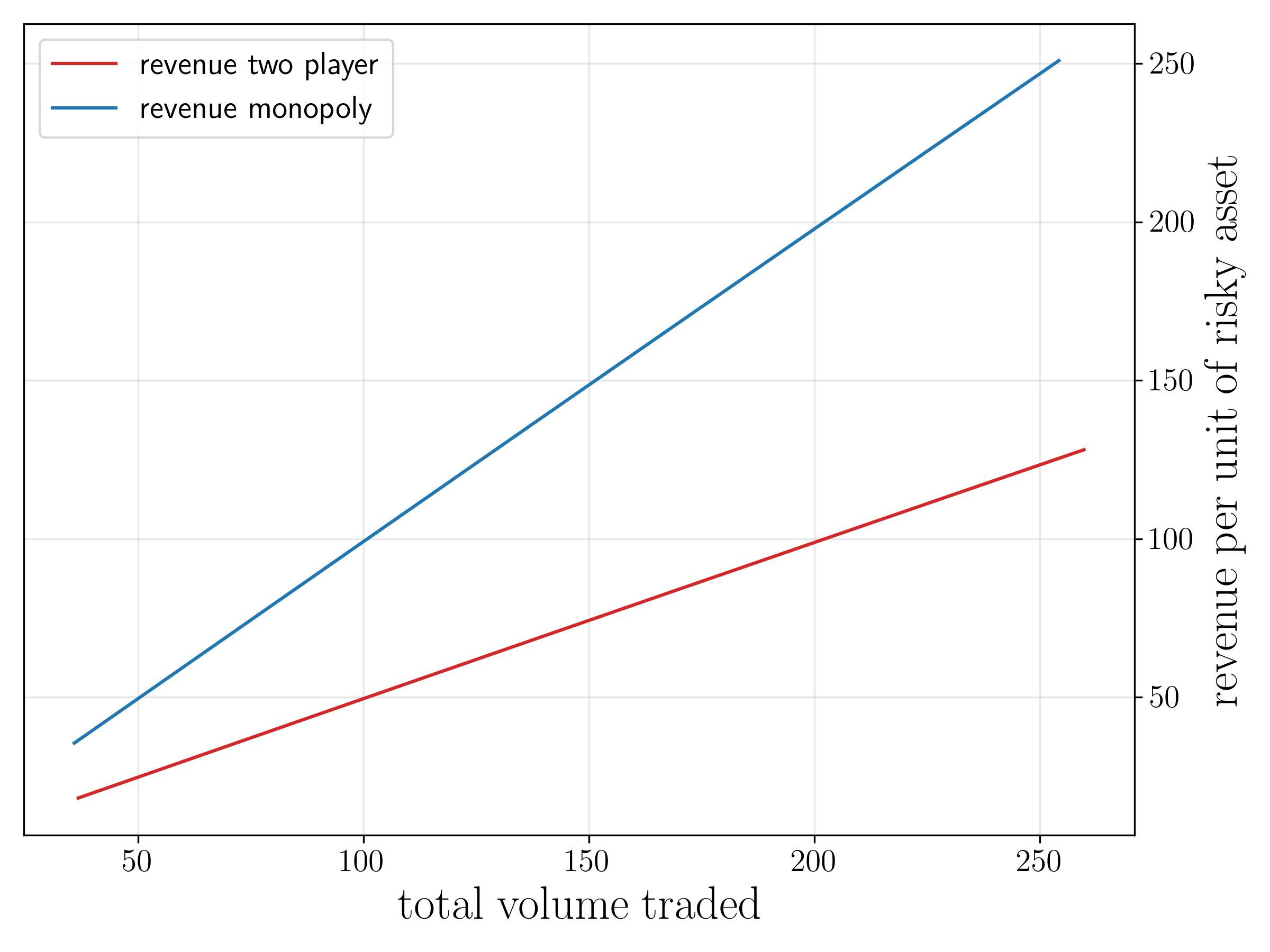}
        \caption{Revenue per player as a function of total traded volume in the monopoly (blue line), two-player (red line) when liquidity is split among all participants.}
        \label{fig:rev per player diff liq}
    \end{subfigure}
\end{figure}

\begin{remark}
    This parametrization is not unique. In particular, one may use an alternative \textit{canonical} calibration that yields a fair comparison with the monopoly benchmark while relying on a different choice of parameters. In this second scenario, the grid for the risky asset $y$ is taken to be identical in the monopoly and duopoly settings. Consequently, the induced grid for the marginal price $Z$ has a step size that is twice as large as in the monopoly case. This is natural: since liquidity is now split across two pools, each pool is shallower, and therefore a trade of a given size produces a larger marginal price impact.

Under this convention, we set $k^{i,0}=k^{i,j}=k/2$, where $k$ denotes the exponential decay parameter in the monopoly case. In addition, let $\lambda^{i,+}=\lambda^{+}/2$ and $\lambda^{i,-}=\lambda^{-}/2$. We tested the numerical results under both parameter conventions and the resulting outputs are of the same order of magnitude. We therefore omit the redundant figures and tables for brevity.\footnote{See the project repository for the corresponding code and additional numerical checks.}
\end{remark}

\section{Conclusion}
In this paper, we studied dynamic fee competition between DEXs and characterised an approximate Nash equilibrium. We found that equilibrium fees alternate between deterring arbitrage and attracting noise-driven trading, but competition shifts the threshold where this change happens from the oracle price to a weighted combination of the oracle and the competing pool's exchange rate. Moreover, we found that these changes benefit strategic liquidity takers while harming liquidity providers; as for noise liquidity takers whether or not they are better off depends on the level of market activity.
\newline
Future research could extend our framework along three main directions. First, by introducing a strategic liquidity provider who jointly chooses fees and an optimal initial liquidity, adjusting provision in response to equilibrium fee schedules. Second, it would be valuable to study the infinite-player limit game, to understand how our alternating-fee equilibrium and the weighted benchmark threshold behave under near-perfect competition. Finally, an important step is to generalize the analysis to concentrated-liquidity AMMs (e.g., Uniswap V3).
\appendix

\section{Multiple-player model}\label{section: M-player model}

We now extend our theory to the case where there are $M$ players. The players will be denoted with a number $i \in \{1, \dots, M \}$.
For each player $i$ the controlled intensities of the processes
$\{ N^{i,-,\feeLTbuys_{t}} \}_{t \in [0,T]}$ and $\{  N^{i,+,\feeLTsells_{t} } \}_{t \in [0,T]}$ are given by
\begin{align*}
\lambda_{t}^{i,-,\fees} & := \lambda^{i,-} \exp{ \left( k^{i,0}((S_{t} - \zeta)  - \exratebuy^{i,\feeLTbuys^{i}}(Y_{t}^{i,\fees}))\Delta_{-}^{i}(Y_{t}^{i,\fees}) + \sum_{\substack{j=1 \\ j\neq i}}^{M} k^{i,j}(\exratebuy^{j}(Y_{t}^{j,\fees}) - \exratebuy^{i,\feeLTbuys^{i}}(Y_{t}^{i,\fees})) \Delta_{-}^{i}(Y_{t}^{i,\fees}) \right) } \ind_{ \{Y_{t}^{i,\fees} > \underline{y}^{i} \}} \\
\lambda_{t}^{i,+,\fees} & := \lambda^{i,+} \exp{ \left( k^{i,0}(\exratesell^{i,\feeLTsells^{i}}(Y_{t}^{i,\fees}) - (S_{t} + \zeta)) \Delta_{+}^{i}(Y_{t}^{i,\fees}) + \sum_{\substack{j=1 \\ j\neq i}}^{M} k^{i,j}(\exratesell^{i,\feeLTsells^{i}}(Y_{t}^{i,\fees}) - \exratesell^{j}(Y_{t}^{j,\fees})) \Delta_{+}^{i}(Y_{t}^{i,\fees}) \right)  } \ind_{ \{Y_{t}^{i,\fees} < \overline{y}^{i} \}}
\end{align*}

The interpretations are the same as in the two-player case. We establish some notation, for each $i \in \{1, \dots, M\}$ we call $\fees^{i} = (\feeLTbuys^{i},\feeLTsells^{i})$, $\feeLTbuys = (\feeLTbuys^{1}, \dots, \feeLTbuys^{M})$ and $\feeLTsells = (\feeLTsells^{1}, \dots, \feeLTsells^{M})$ $\fees = (\fees^{1}, \dots \fees^{M})$.

\medskip{}
For each $i \in \{1, \dots, M\}$ the cumulative fees $\{\Cash_{t}^{i,\fees^{i}}\}_{t \in [0, T]}$ collected by the pool $i$ are in turn given by
\begin{equation*} \label{eq : SDE for holdings M players}
\Cash_{t}^{i,\fees^{i}} := \int_0^t \feeLTsells_{u}^{i} \exratesell^{i,\feeLTsells_{u}^{i}}(Y_{u}^{i,\fees}) \Delta_{+}^{i}(Y_{t}^{i,\fees}) \dd N^{i,+,\feeLTsells^{i}}_{u} + \int_0^t \feeLTbuys^{i}_{u}\exratebuy^{i,\feeLTbuys^{i}_{u}}(Y_{u}^{i,\fees}) \Delta_{-}^{i}(Y_{t}^{i,\fees}) \dd N^{i,+,\feeLTbuys^{i}}_{u}, \quad \quad t \in [0,T],
\end{equation*}

Each pool seeks to solve the control problems
\begin{align*}
    \omega^{i}(t,s,\cash^{i},y^{1}, \dots, y^{M}) & : = \sup_{\fees^{i} \in \mathcal{A}_{t}} \omega^{(i,\fees^{i})}(t,s,\cash^{i},y^{1}, \dots, y^{M})
\end{align*}
where $\mathcal{A}_{t}$ denotes the set of all $\mathbb{F}$-predictable and bounded fee structure processes $(\feeLTbuys_{u}^{i},\feeLTsells_{u}^{i})_{\{t \leq u \leq T\}}$ and the conditional performance criterion are given by \[ \omega^{i}(t,s,\cash^{i},y^{1}, \dots, y^{M}) : = \mathbb{E}_{(t,s,\cash^{a},y^{1}, \dots, y^{M})} \left[ \Cash_{T}^{(i,t,s,\cash^{i},y^{1}, \dots, y^{M},\fees^{i})} \right], \quad i \in \{1, \dots, M\} \]

Here, for each \(i\in\{1,\dots,M\}\),
\[
\big\{\Cash_{u}^{(i,t,s,\cash^{i},y^{1},\dots,y^{M},\fees^{i})}\big\}_{u\in[t,T]},\quad
\big\{Y_{u}^{(i,t,s,\cash^{i},y^{1},\dots,y^{M},\fees^{i})}\big\}_{u\in[t,T]},\quad
\big\{S_{u}^{(i,t,s,\cash^{i},y^{1},\dots,y^{M})}\big\}_{u\in[t,T]}
\]
denote the (controlled) processes \(\Cash^{i}\), \(Y^{i}\), and \(S\) restarted at time \(t\) with initial values \(\cash^{i}\), \(y^{1},\dots,y^{M}\), and \(s\), respectively. Clearly all the value functions are linear in $\cash^{i}$ and we denote with $v^{i}$ the value function after the linear ansatz for $\cash^{i}$. From the dynamic programming principle, we determine that the Hamilton-Jacobi-Bellman (HJB) for the value function of the player $i$ is

\small
\begin{align*}
0 \;=\;&
\frac{\partial}{\partial t}\, v^{i}(t,s,y^{1},\dots,y^{M})
\;+\; \frac{\sigma^{2}}{2}\,\frac{\partial^{2}}{\partial s^{2}} v^{i}(t,s,y^{1},\dots,y^{M}) \\[4pt]
&\;+\; \sup_{\feeLTbuys^{i}\in\mathbb{R}}
\lambda^{i,-}\, e^{\,k^{i}\!\left(\tfrac{k^{i,0}}{k^{i}}s
+ \sum_{\substack{j=1\\ j\neq i}}^{M}\tfrac{k^{i,j}}{k^{i}}\,\exratebuy^{j}(y^{j})
- (1+\feeLTbuys^{i})\,\exratebuy^{i}(y^{i})\right)\Delta_-^{i}(y^{i})}\,
\Big( v^{i}(t,s,y^{i,-})-v^{i}(t,s,y) + \feeLTbuys^{i}\,\exratebuy^{i}(y^{i})\,\Delta_{-}^{i}(y^{i}) \Big)\,
\ind_{\{y^{i}>\underline y^{i}\}} \\[6pt]
&\;+\; \sup_{\feeLTsells^{i}\in\mathbb{R}}
\lambda^{i,+}\, e^{\,k^{i}\!\left( (1-\feeLTsells^{i})\,\exratesell^{i}(y^{i})
- \tfrac{k^{i,0}}{k^{i}}s
- \sum_{\substack{j=1\\ j\neq i}}^{M}\tfrac{k^{i,j}}{k^{i}}\,\exratesell^{j}(y^{j})\right)\Delta_+^{i}(y^{i})}\,
\Big( v^{i}(t,s,y^{i,+})-v^{i}(t,s,y) + \feeLTsells^{i}\,\exratesell^{i}(y^{i})\,\Delta_{+}^{i}(y^{i}) \Big)\,
\ind_{\{y^{i}<\overline y^{i}\}} \\[6pt]
&\;+\; \sum_{\substack{j=1\\ j\neq i}}^{M}
\lambda^{j,-}\, e^{\,k^{j}\!\left(\tfrac{k^{j,0}}{k^{j}}s
+ \sum_{\substack{\ell=1\\ \ell\neq j}}^{M}\tfrac{k^{j,\ell}}{k^{j}}\,\exratebuy^{\ell}(y^{\ell})
- (1+\feeLTbuys^{j})\,\exratebuy^{j}(y^{j})\right)\Delta_-^{j}(y^{j})}\,
\Big( v^{i}(t,s,y^{j,-})-v^{i}(t,s,y) \Big)\,
\ind_{\{y^{j}>\underline y^{j}\}} \\[6pt]
&\;+\; \sum_{\substack{j=1\\ j\neq i}}^{M}
\lambda^{j,+}\, e^{\,k^{j}\!\left( (1-\feeLTsells^{j})\,\exratesell^{j}(y^{j})
- \tfrac{k^{j,0}}{k^{j}}s
- \sum_{\substack{\ell=1\\ \ell\neq j}}^{M}\tfrac{k^{j,\ell}}{k^{j}}\,\exratesell^{\ell}(y^{\ell})\right)\Delta_+^{j}(y^{j})}\,
\Big( v^{i}(t,s,y^{j,+})-v^{i}(t,s,y) \Big)\,
\ind_{\{y^{j}<\overline y^{j}\}}.
\end{align*}
\normalsize

\noindent
Here \(k^{i}:=k^{i,0}+\sum_{\substack{j=1\\ j\neq i}}^{M}k^{i,j}\) for each \(i\in\{1,\dots,M\}\), and we use the shorthand
\[
y^{j,-} := (y^{1},\dots,y^{j}-\Delta_{-}^{j}(y^{j}),\dots,y^{M}),\qquad
y^{j,+} := (y^{1},\dots,y^{j}+\Delta_{+}^{j}(y^{j}),\dots,y^{M}).
\]
The terminal condition is \(v^{i}(T,s,y^{1},\dots,y^{M})=0\).
\newline
First order condition on the maximizer yields, for every $i \in \{1, \dots M\}$,

\small
\begin{align*}
\feeLTbuys^{i,*}(t,s,y^{1},\dots,y^{M})
&= \frac{1 + k^{i}\big(v^{i}(t,s,y^{1},\dots,y^{M}) - v^{i}(t,s,y^{i,-})\big)}
{k^{i}\,\exratebuy^{i}(y^{i})\,\Delta_{-}^{i}(y^{i})},\\[4pt]
\feeLTsells^{i,*}(t,s,y^{1},\dots,y^{M})
&= \frac{1 + k^{i}\big(v^{i}(t,s,y^{1},\dots,y^{M}) - v^{i}(t,s,y^{i,+})\big)}
{k^{i}\,\exratesell^{i}(y^{i})\,\Delta_{+}^{i}(y^{i})}.
\end{align*}
\normalsize

\noindent
plugging back into the original equation yields

\small
\begin{align*}
0 \;=\;& \frac{\partial}{\partial t} v^{i}(t,s,y^{1},\dots,y^{M})
\;+\; \frac{\sigma^{2}}{2}\,\frac{\partial^{2}}{\partial s^{2}} v^{i}(t,s,y^{1},\dots,y^{M}) \\[6pt]
&\;+\; \frac{\lambda^{i,-}\,
e^{\,k^{i}\!\left(\tfrac{k^{i,0}}{k^{i}} s + \sum_{\substack{j=1\\ j\neq i}}^{M}\tfrac{k^{i,j}}{k^{i}}\,\exratebuy^{j}(y^{j})\right)\Delta_{-}^{i}(y^{i}) - 1}}
{k^{i}}\;
e^{-\,k^{i}\,\exratebuy^{i}(y^{i})\,\Delta_{-}^{i}(y^{i})}\;
e^{\,k^{i}\,\big(v^{i}(t,s,y^{i,-}) - v^{i}(t,s,y)\big)}\;
\ind_{\{y^{i}>\underline y^{i}\}} \\[6pt]
&\;+\; \sum_{\substack{j=1\\ j\neq i}}^{M}
\lambda^{j,-}\,
e^{\,k^{j}\!\left(\tfrac{k^{j,0}}{k^{j}} s + \sum_{\substack{\ell=1\\ \ell\neq j}}^{M}\tfrac{k^{j,\ell}}{k^{j}}\,\exratebuy^{\ell}(y^{\ell})\right)\Delta_{-}^{j}(y^{j}) - 1}\;
e^{-\,k^{j}\,\exratebuy^{j}(y^{j})\,\Delta_{-}^{j}(y^{j})}\;
e^{\,k^{j}\,\big(v^{j}(t,s,y^{j,-}) - v^{j}(t,s,y)\big)} \\
&\hspace{5em}\times \big(v^{i}(t,s,y^{j,-}) - v^{i}(t,s,y)\big)\;
\ind_{\{y^{j}>\underline y^{j}\}} \\[6pt]
&\;+\; \frac{\lambda^{i,+}\,
e^{-\,k^{i}\!\left(\tfrac{k^{i,0}}{k^{i}} s + \sum_{\substack{j=1\\ j\neq i}}^{M}\tfrac{k^{i,j}}{k^{i}}\,\exratesell^{j}(y^{j})\right)\Delta_{+}^{i}(y^{i}) - 1}}
{k^{i}}\;
e^{\,k^{i}\,\exratesell^{i}(y^{i})\,\Delta_{+}^{i}(y^{i})}\;
e^{\,k^{i}\,\big(v^{i}(t,s,y^{i,+}) - v^{i}(t,s,y)\big)}\;
\ind_{\{y^{i}<\overline y^{i}\}} \\[6pt]
&\;+\; \sum_{\substack{j=1\\ j\neq i}}^{M}
\lambda^{j,+}\,
e^{-\,k^{j}\!\left(\tfrac{k^{j,0}}{k^{j}} s + \sum_{\substack{\ell=1\\ \ell\neq j}}^{M}\tfrac{k^{j,\ell}}{k^{j}}\,\exratesell^{\ell}(y^{\ell})\right)\Delta_{+}^{j}(y^{j}) - 1}\;
e^{\,k^{j}\,\exratesell^{j}(y^{j})\,\Delta_{+}^{j}(y^{j})}\;
e^{\,k^{j}\,\big(v^{j}(t,s,y^{j,+}) - v^{j}(t,s,y)\big)} \\
&\hspace{5em}\times \big(v^{i}(t,s,y^{j,+}) - v^{i}(t,s,y)\big)\;
\ind_{\{y^{j}<\overline y^{j}\}},
\end{align*}
\normalsize

\noindent
where \(k^{i}:=k^{i,0}+\sum_{\substack{j=1\\ j\neq i}}^{M}k^{i,j}\) for each \(i=1,\dots,M\). We now make a few assumptions to compute an approximate solution. First, we treat the CEX price $S_{t}$ as a parameter. Second, we assume that a change in one agent’s inventory has a negligible effect on any other agent’s value function at first order.
For a fixed player \(i\in\{1,\dots,M\}\) and any \(j\neq i\),
\begin{align*}
v^{i}\bigl(t, y^{1},\dots, y^{j}+\Delta_{+}^{j}(y^{j}),\dots, y^{M}\bigr)
&\approx v^{i}\bigl(t, y^{1},\dots, y^{M}\bigr),\\
v^{i}\bigl(t, y^{1},\dots, y^{j}-\Delta_{-}^{j}(y^{j}),\dots, y^{M}\bigr)
&\approx v^{i}\bigl(t, y^{1},\dots, y^{M}\bigr).
\end{align*}
Under these assumptions, \(v^{i}\) no longer depends on \(s\) and we write \(v^{i}(t,y^{1},\dots,y^{M})\).
Plugging the maximizers and the HJB for player \(i\) reduces to
\small
\begin{align}
0
&=
\frac{\lambda^{i,-}}{k^{i}}
\exp\!\Big(
k^{i}\Big(\tfrac{k^{i,0}}{k^{i}}\,s
+ \sum_{\substack{j=1\\ j\neq i}}^{M}\tfrac{k^{i,j}}{k^{i}}\,\exratebuy^{j}(y^{j})\Big)
\Delta_{-}^{i}(y^{i}) - 1
\Big)\,
\exp\!\big(-k^{i}\exratebuy^{i}(y^{i})\Delta_{-}^{i}(y^{i})\big) \notag\\
&\quad\times
\exp\!\Big(
k^{i}\big[
v^{i}(t, y^{1},\dots, y^{i}-\Delta_{-}^{i}(y^{i}), \dots, y^{M})
- v^{i}(t, y^{1},\dots, y^{M})
\big]
\Big)\,
\ind_{\{y^{i}>\underline y^{i}\}}
\notag\\[4pt]
&\;+\;
\frac{\lambda^{i,+}}{k^{i}}
\exp\!\Big(
-\,k^{i}\Big(\tfrac{k^{i,0}}{k^{i}}\,s
+ \sum_{\substack{j=1\\ j\neq i}}^{M}\tfrac{k^{i,j}}{k^{i}}\,\exratesell^{j}(y^{j})\Big)
\Delta_{+}^{i}(y^{i}) - 1
\Big)\,
\exp\!\big(k^{i}\exratesell^{i}(y^{i})\Delta_{+}^{i}(y^{i})\big) \notag\\
&\quad\times
\exp\!\Big(
k^{i}\big[
v^{i}(t, y^{1},\dots, y^{i}+\Delta_{+}^{i}(y^{i}), \dots, y^{M})
- v^{i}(t, y^{1},\dots, y^{M})
\big]
\Big)\,
\ind_{\{y^{i}<\overline y^{i}\}}
\notag\\[4pt]
&\;+\; \frac{\partial}{\partial t}\, v^{i}(t, y^{1},\dots, y^{M}).
\end{align}
\normalsize

\noindent

we introduce, for each \(i\in\{1,\dots,M\}\),
\[
e^{\,k^{i} v^{i}(t,y^{1},\dots,y^{M})} \;:=\; w^{i}(t,y^{1},\dots,y^{M}),
\qquad\text{so that}\qquad
w^{i}(T,y^{1},\dots,y^{M})=1.
\]

\noindent
With this change of variables, the HJB equation for player \(i\) becomes linear:
\small
\begin{align*}
0 \;=\;&
\frac{\partial}{\partial t}\, w^{i}\!\bigl(t, y^{1},\dots,y^{M}\bigr)
\\[4pt]
&\;+\;
\lambda^{i,-}\;
\exp\!\Biggl(
k^{i}\!\left(\frac{k^{i,0}}{k^{i}}\,S_{0}
+ \sum_{\substack{j=1\\ j\neq i}}^{M}\frac{k^{i,j}}{k^{i}}\,\exratebuy^{j}(y^{j})\right)\!\Delta_{-}^{i}(y^{i}) - 1
\Biggr)\,
\exp\!\Bigl(-k^{i}\,\exratebuy^{i}(y^{i})\,\Delta_{-}^{i}(y^{i})\Bigr)
\\
&\qquad\times\;
w^{i}\!\bigl(t, y^{1},\dots, y^{i}-\Delta_{-}^{i}(y^{i}), \dots, y^{M}\bigr)\;
\ind_{\{y^{i}>\underline y^{i}\}}
\\[8pt]
&\;+\;
\lambda^{i,+}\;
\exp\!\Biggl(
-\,k^{i}\!\left(\frac{k^{i,0}}{k^{i}}\,S_{0}
+ \sum_{\substack{j=1\\ j\neq i}}^{M}\frac{k^{i,j}}{k^{i}}\,\exratesell^{j}(y^{j})\right)\!\Delta_{+}^{i}(y^{i}) - 1
\Biggr)\,
\exp\!\Bigl(+k^{i}\,\exratesell^{i}(y^{i})\,\Delta_{+}^{i}(y^{i})\Bigr)
\\
&\qquad\times\;
w^{i}\!\bigl(t, y^{1},\dots, y^{i}+\Delta_{+}^{i}(y^{i}), \dots, y^{M}\bigr)\;
\ind_{\{y^{i}<\overline y^{i}\}}.
\end{align*}
\normalsize

The formal result is summarized in the following theorem

\begin{theorem} \label{th: nash equilibrium multiplayer theorem}
    Fix an index $h \in \{1, \dots , M\}$ and a multindex $l \in \bigotimes_{\substack{j=1\\ j\neq i}}^{M} \{-N^{j}, \dots, N^{j} \}$. Denote with $\mathbf{y}^{l}$ the vector \[\mathbf{y}^{l} : = (y^{1,l^{1}}, \dots,y^{i-1,l^{i-1}},y^{i+1,l^{i+1}} ,\dots,y^{M,l^{M}} ).  \] Define the matrix $ \mathbf{A}^{h,l} : =  ( \mathbf{A}^{l}_{i,j} )_{0 \leq i \leq j \leq 2N^{i}}$ by 
    \small
\begin{align*}
\mathbf{A}^{h}_{i,j}(\mathbf{y}^{l})
\;:=\;
\left\{
\begin{array}{@{}l@{}}
\lambda^{i,+}\,
\exp\!\Bigg(
-\,k^{h}\!\Big(\frac{k^{h,0}}{k^{h}}\,S_{0}
+ \sum_{\substack{m=1\\ m\neq i}}^{M}\frac{k^{h,m}}{k^{h}}\,\exratesell^{m}(y^{m,l^{m}})\Big)\,
\Delta_{+}^{h}(y^{h,j-N^{h}}) - 1
\Bigg)
\\
\qquad\times
\exp\!\Big(
k^{h}\,\exratesell^{h}(y^{h,j-N^{h}})\,\Delta_{+}^{h}(y^{h,j-N^{h}})
\Big),
\quad \text{if } i=j-1,
\\[8pt]
\lambda^{i,-}\,
\exp\!\Bigg(
\;\;k^{h}\!\Big(\frac{k^{h,0}}{k^{h}}\,S_{0}
+ \sum_{\substack{m=1\\ m\neq i}}^{M}\frac{k^{h,m}}{k^{h}}\,\exratebuy^{m}(y^{m,l^{m}})\Big)\,
\Delta_{-}^{h}(y^{h,j-N^{h}}) - 1
\Bigg)
\\
\qquad\times
\exp\!\Big(
-\,k^{h}\,\exratebuy^{h}(y^{h,j-N^{h}})\,\Delta_{-}^{h}(y^{h,j-N^{h}})
\Big),
\quad \text{if } i=j+1,
\\[6pt]
0,\quad \text{otherwise.}
\end{array}
\right.
\end{align*}
Denote with $\mathbf{1}$ the unit vectors of $\mathbb{R}^{2N^{h} + 1}$. Define the functions $w^{h}: [0,T] \bigotimes_{\substack{j=1}}^{M} \{y^{j,-N^{j}}, \dots, y^{j,N^{j}} \} \to \mathbb{R}$ by
\begin{align*}
    w^{h}(t,y^{h,i},\mathbf{y}^{l}) &:= \left( \exp\big( \mathbf{A}^{h}(\mathbf{y}^{l})(T - t) \big) \, \mathbf{1} \right)_{i}
\end{align*}
and the functions $v^{h}: [0,T] \bigotimes_{\substack{j=1}}^{M} \{y^{j,-N^{j}}, \dots, y^{j,N^{j}} \} \times \mathbb{R}_{+} \to \mathbb{R}$, $v^{b}: [0,T] \times \{ y^{a,-N^{a}}, \dots, y^{a,N^{a}} \} \times \{ y^{b,-N^{b}}, \dots, y^{b,N^{b}} \} \times \mathbb{R}_{+} \to \mathbb{R}$ as 
\begin{align*}
    v^{h}(t,y^{h,i},\mathbf{y}^{l}, \cash^{h}) & := \cash^{h} + \frac{1}{k^{h}} \log(w^{h}(t,y^{h,i},\mathbf{y}^{l}).
\end{align*}
Then the functions $v^{h}$ solve the system of HJBs

\begin{align*}
\begin{cases}
0 & \;=\;
\frac{\partial}{\partial t}\, v^{h}(t,y^{h,i},\mathbf{y}^{l})
\;+\; \frac{\sigma^{2}}{2}\,\frac{\partial^{2}}{\partial s^{2}} v^{h}(t,y^{h,i},\mathbf{y}^{l}) \\ & \;+\; \sup_{\feeLTbuys^{h}\in\mathbb{R}} \lambda^{h,-}\,
\exp\Bigg\{ k^{h}\!\Big( \tfrac{k^{h,0}}{k^{h}}\,s
+ \sum_{\substack{j=1 j\neq h}}^{M}\tfrac{k^{h,j}}{k^{h}}\,\exratebuy^{j}(y^{j}) \Big) \\
& \quad - k^{h}\!\Big(1+\feeLTbuys^{h}\Big)\exratebuy^{h}(y^{h}) \,\Delta_-^{h}(y^{h,i}) \Bigg\} \Big( v^{h}(t, y^{h,i} - \Delta_-^{h}(y^{h,i}), \mathbf{y}^{l})
      - v^{h}(t, y^{h,i}, \mathbf{y}^{l})
      + \feeLTbuys^{h}\,\exratebuy^{h}(y^{h})\,\Delta_-^{h}(y^{h,i}) \Big)\,
\ind_{\{y^{h,i}>\underline y^{h}\}} \\[6pt]
&\;+\; \sup_{\feeLTsells^{h}\in\mathbb{R}} \lambda^{h,+}\,
\exp\Bigg\{ -k^{h}\!\Big( \tfrac{k^{h,0}}{k^{h}}\,s
+ \sum_{\substack{j=1\\ j\neq h}}^{M}\tfrac{k^{h,j}}{k^{h}}\,\exratesell^{j}(y^{j}) \Big) \\
&\quad
+ k^{h}\!\Big(1-\feeLTsells^{h}\Big)\exratesell^{h}(y^{h}) \,\Delta_+^{h}(y^{h,i}) \Bigg\}
\Big( v^{h}(t, y^{h,i} + \Delta_+^{h}(y^{h,i}), \mathbf{y}^{l})
      - v^{h}(t, y^{h,i}, \mathbf{y}^{l})
      + \feeLTsells^{h}\,\exratesell^{h}(y^{h})\,\Delta_+^{h}(y^{h,i}) \Big)\,
\ind_{\{y^{h,i} < \overline y^{h}\}} \\[6pt]
\end{cases}
\end{align*}
with boundary conditions $v^{h}(T,y^{h,i},\mathbf{y}^{l},\cash^{h}) = \cash^{h}$ for every $h$, $i$ and $l$. Moreover, the corresponding maximizers are independent of $\cash^{h}$ and are given by
\begin{align*}
\feeLTbuys^{h,*}(t,y^{h,i},\mathbf{y}^{l})
&= \frac{1 + k^{h}\big(v^{h}(t,y^{h,i},\mathbf{y}^{l}) - v^{h}(t,y^{h,i} - \Delta_{-}^{h}(y^{h,i}),\mathbf{y}^{l})\big)}
{k^{h}\,\exratebuy^{h}(y^{h,i})\,\Delta_{-}^{h}(y^{h,i})},\\[4pt]
\feeLTsells^{h,*}(t,y^{h,i},\mathbf{y}^{l})
&= \frac{1 + k^{h}\big(v^{h}(t,y^{h,i},\mathbf{y}^{l}) - v^{h}(t,y^{h,i} + \Delta_{+}^{h}(y^{h,i}),\mathbf{y}^{l})\big)}{k^{h}\,\exratesell^{h}(y^{h,i})\,\Delta_{+}^{h}(y^{h,i})}.
\end{align*}

\end{theorem}

\section{Simulations}\label{section: M-player model simulations}

As in Section \ref{section: two player model simulations} we now perform numerical simulations in a three-player setting and compare the results with the duopoly and monopoly case. The total depth is split between three pools, and for simplicity we assume complete symmetry. In order to make the comparison meaningful we assume that the total depth present in the market is the same as in the monopoly and duopoly scenario.
\begin{table}[H]
\centering
\begin{tabular}{@{}cccccccccc@{}}
\toprule
& & \multicolumn{4}{c}{$\lambda = 100$} & \multicolumn{4}{c}{$\lambda = 150$} \\
\cmidrule(lr){3-6} \cmidrule(lr){7-10}
Player & Type & Fees & Sell & Buy & Vol & Fees & Sell & Buy & Vol \\
\midrule
\multirow{3}{*}{Player 1}  
 & Optimal  & 12.25 & 37.09 & 37.08 & 1236.21 & 18.24 & 55.22 & 55.14 & 1839.57 \\
 & Linear   & 12.25 & 37.06 & 37.11 & 1236.09 & 18.23 & 55.16 & 55.21 & 1839.43 \\
 & Constant & 12.09 & 37.75 & 37.77 & 1258.69 & 17.95 & 56.24 & 56.25 & 1874.91 \\
\cmidrule(lr){1-10}
\multirow{3}{*}{Player 2}  
 & Optimal  & 12.26 & 37.11 & 37.09 & 1236.69 & 18.24 & 55.24 & 55.17 & 1840.33 \\
 & Linear   & 12.26 & 37.07 & 37.12 & 1236.59 & 18.24 & 55.18 & 55.23 & 1840.25 \\
 & Constant & 12.09 & 37.75 & 37.78 & 1258.84 & 17.95 & 56.24 & 56.26 & 1875.11 \\
\cmidrule(lr){1-10}
\multirow{3}{*}{Player 3}  
 & Optimal  & 12.25 & 37.09 & 37.08 & 1236.10 & 18.24 & 55.22 & 55.16 & 1839.83 \\
 & Linear   & 12.25 & 37.05 & 37.11 & 1235.99 & 18.24 & 55.16 & 55.23 & 1839.74 \\
 & Constant & 12.09 & 37.73 & 37.76 & 1258.27 & 17.95 & 56.22 & 56.25 & 1874.57 \\
\cmidrule(lr){1-10}
\multirow{3}{*}{Total}  
 & Optimal  & 36.76 & 111.29 & 111.25 & 3709.00 & 54.72 & 165.68 & 165.47 & 5519.73 \\
 & Linear   & 36.76 & 111.18 & 111.34 & 3708.67 & 54.71 & 165.50 & 165.67 & 5519.42 \\
 & Constant & 36.27 & 113.23 & 113.31 & 3775.80 & 53.85 & 168.70 & 168.76 & 5624.59 \\
\bottomrule
\end{tabular}
\caption{Average fees collected (Fees), number of sell trades (Sell), number of buy trades (Buy), and volume traded (Vol) for Players 1,2,3 and Total under different fee types when $k=2$, with $\lambda\in\{100,150\}$.}
\label{tab:simulations_3players_k2}
\end{table}

\begin{table}[H]
\centering
\begin{tabular}{@{}cccccccccc@{}}
\toprule
& & \multicolumn{4}{c}{$\lambda = 100$} & \multicolumn{4}{c}{$\lambda = 150$} \\
\cmidrule(lr){3-6} \cmidrule(lr){7-10}
Player & Type & Fees & Sell & Buy & Vol & Fees & Sell & Buy & Vol \\
\midrule
\multirow{3}{*}{Player 1}  
 & Optimal  & 24.22 & 36.51 & 36.52 & 1217.23 & 36.04 & 54.35 & 54.31 & 1810.96 \\
 & Linear   & 24.22 & 36.49 & 36.54 & 1217.20 & 36.04 & 54.31 & 54.35 & 1810.94 \\
 & Constant & 24.12 & 36.76 & 36.80 & 1225.93 & 35.84 & 54.76 & 54.78 & 1825.60 \\
\cmidrule(lr){1-10}
\multirow{3}{*}{Player 2}  
 & Optimal  & 24.23 & 36.52 & 36.54 & 1217.71 & 36.06 & 54.36 & 54.34 & 1811.78 \\
 & Linear   & 24.23 & 36.50 & 36.56 & 1217.69 & 36.05 & 54.32 & 54.38 & 1811.78 \\
 & Constant & 24.13 & 36.77 & 36.81 & 1226.25 & 35.85 & 54.77 & 54.81 & 1826.21 \\
\cmidrule(lr){1-10}
\multirow{3}{*}{Player 3}  
 & Optimal  & 24.22 & 36.51 & 36.52 & 1217.18 & 36.04 & 54.34 & 54.33 & 1811.14 \\
 & Linear   & 24.22 & 36.49 & 36.54 & 1217.17 & 36.04 & 54.30 & 54.37 & 1811.18 \\
 & Constant & 24.12 & 36.75 & 36.79 & 1225.71 & 35.84 & 54.74 & 54.79 & 1825.58 \\
\cmidrule(lr){1-10}
\multirow{3}{*}{Total}  
 & Optimal  & 72.67 & 109.54 & 109.58 & 3652.12 & 108.14 & 163.05 & 162.98 & 5433.88 \\
 & Linear   & 72.67 & 109.48 & 109.64 & 3652.06 & 108.13 & 162.93 & 163.10 & 5433.90 \\
 & Constant & 72.37 & 110.28 & 110.40 & 3677.89 & 107.53 & 164.27 & 164.38 & 5477.39 \\
\bottomrule
\end{tabular}
\caption{Average fees collected (Fees), number of sell trades (Sell), number of buy trades (Buy), and volume traded (Vol) for Players 1,2,3 and Total under different fee types when $k=1$, with $\lambda\in\{100,150\}$.}
\label{tab:simulations_3players_k1}
\end{table}

We perform analogous simulations in the three-player case and obtain qualitatively similar, but amplified, effects. First, increased competition further benefits a strategic liquidity taker: with three venues, the liquidity taker can route orders to the best available quotes and thus extract more value than in the duopoly case. Second, in low-volatility regimes this improvement is largely paid for by noise liquidity takers, who face a higher average slippage as competition intensifies. As volatility rises, however, this pattern reverses: the average slippage tends to decrease with the number of competing venues, indicating that in high-volatility regimes liquidity takers on average obtain better execution when more players compete for order flow.

\begin{figure}[H]
    \centering
    \begin{subfigure}[t]{0.48\textwidth}
        \centering
        \includegraphics[width=\textwidth]{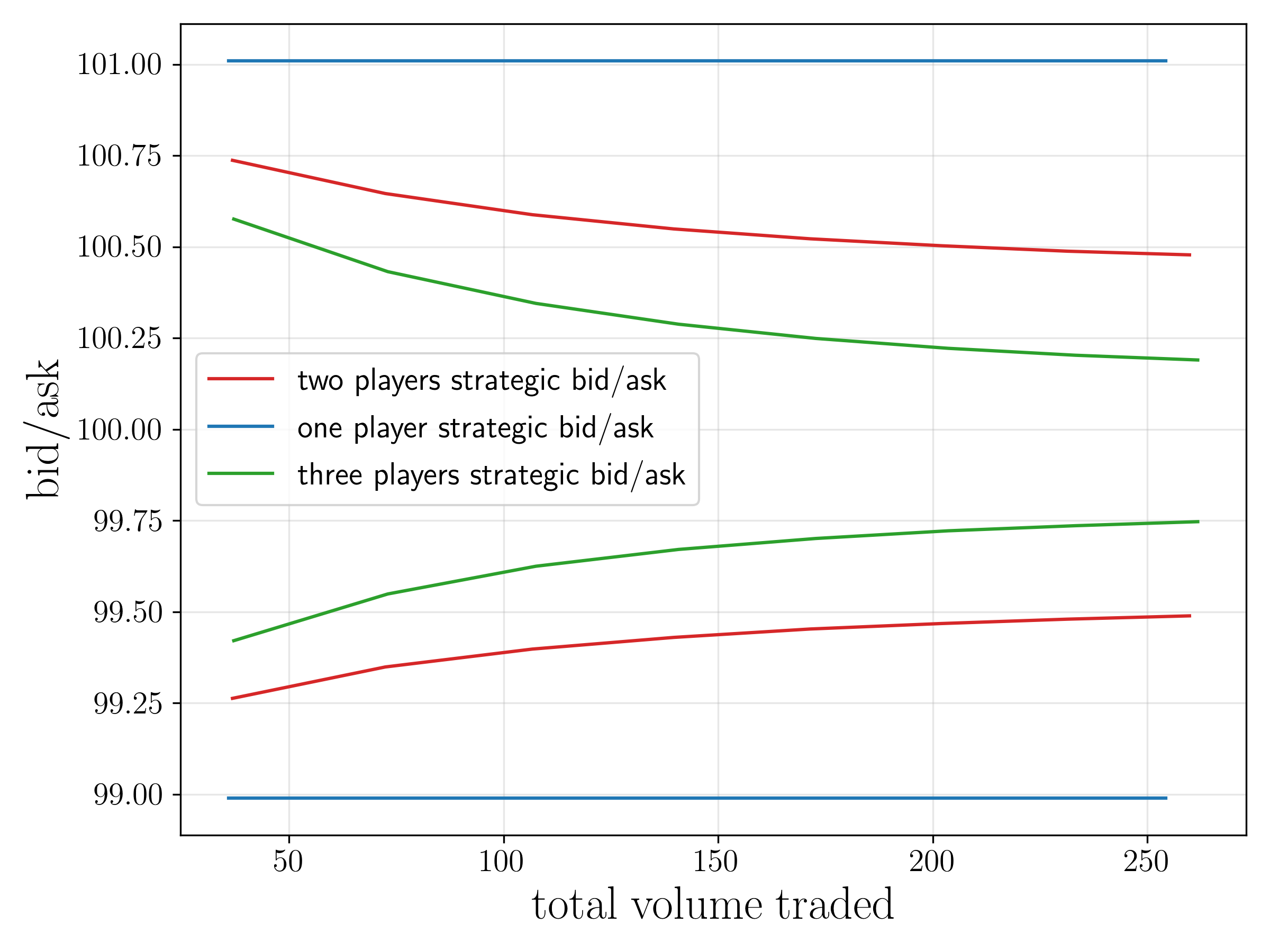}
        \caption{Strategic Bid--ask as a function of total traded volume in the monopoly (blue line), two players (red line) and three players (green line) case.}
        \label{fig:bid-ask-3players}
    \end{subfigure}
    \hspace*{0.01\textwidth}
    \begin{subfigure}[t]{0.48\textwidth}
        \centering
        \includegraphics[width=\textwidth]{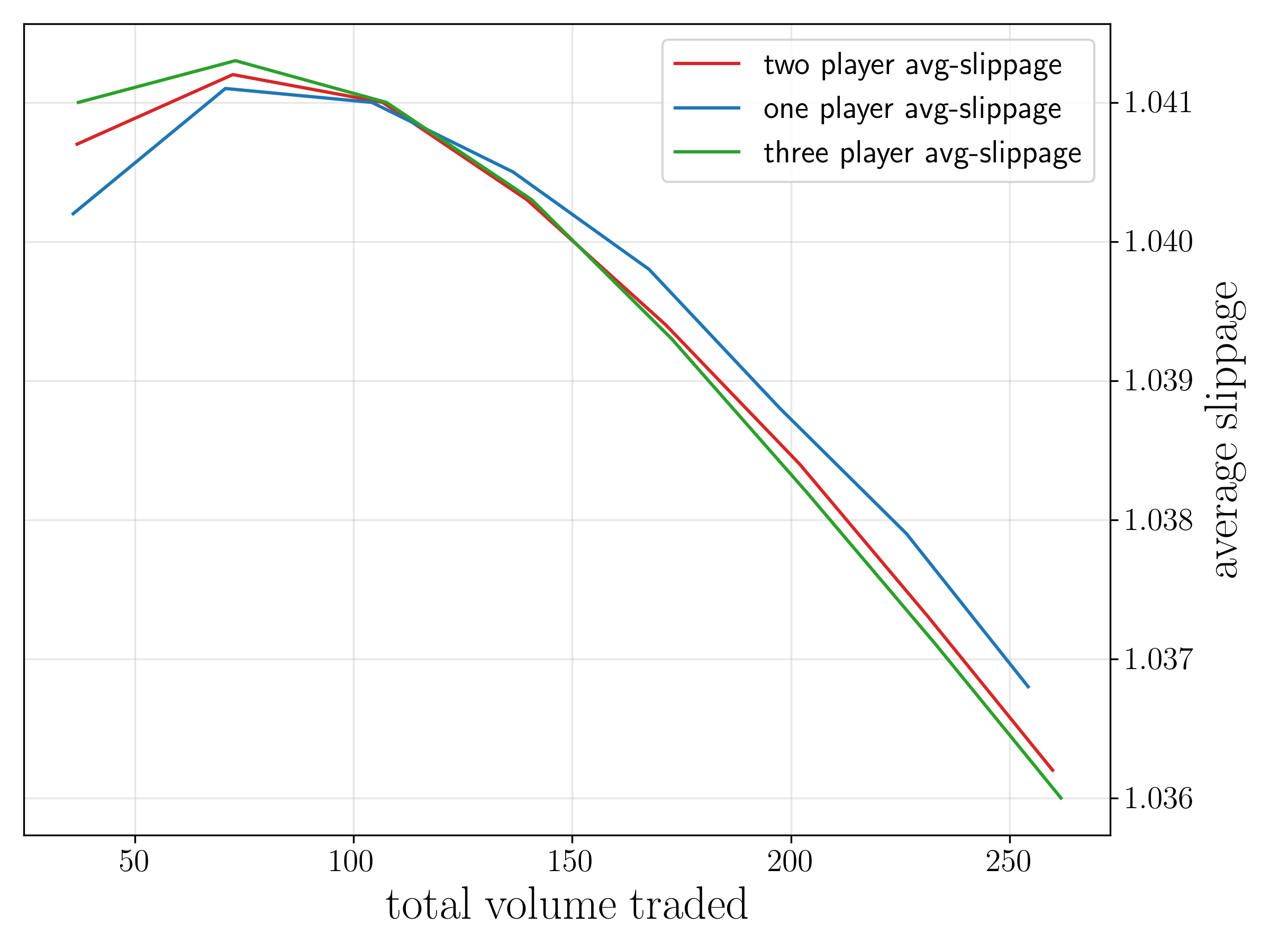}
        \caption{Average slippage as a function of total traded volume in the monopoly (blue line), two players (red line) and three players (green line) case.}
        \label{fig:slippage-3players}
    \end{subfigure}
\end{figure}

Finally, we plot the revenue per player and the AMM’s revenue in two cases. In the first case (left figure), we show the revenue of a single player in the monopoly (blue line), two-player (red line), and three-player (green line) scenarios, under the assumption that liquidity is split among all participants. We observe that the revenue per player decreases as the number of players increases. If the number of players continues to grow, the revenue for an individual player may eventually become too low to make participation profitable. This suggests that if an LP must pay an entrance fee to create its own pool, it will only choose to enter if the expected revenue exceeds the entrance fee, and therefore only if the number of players is not too large.

In the second case (right figure), we plot the AMM’s total revenue in a setting where all players provide the same amount of liquidity. Here, we see that total revenue decreases as the number of players increases, because LPs trade at better prices when there are more players.

\begin{figure}[H]
    \centering
    \begin{subfigure}[t]{0.48\textwidth}
        \centering
        \includegraphics[width=\textwidth]{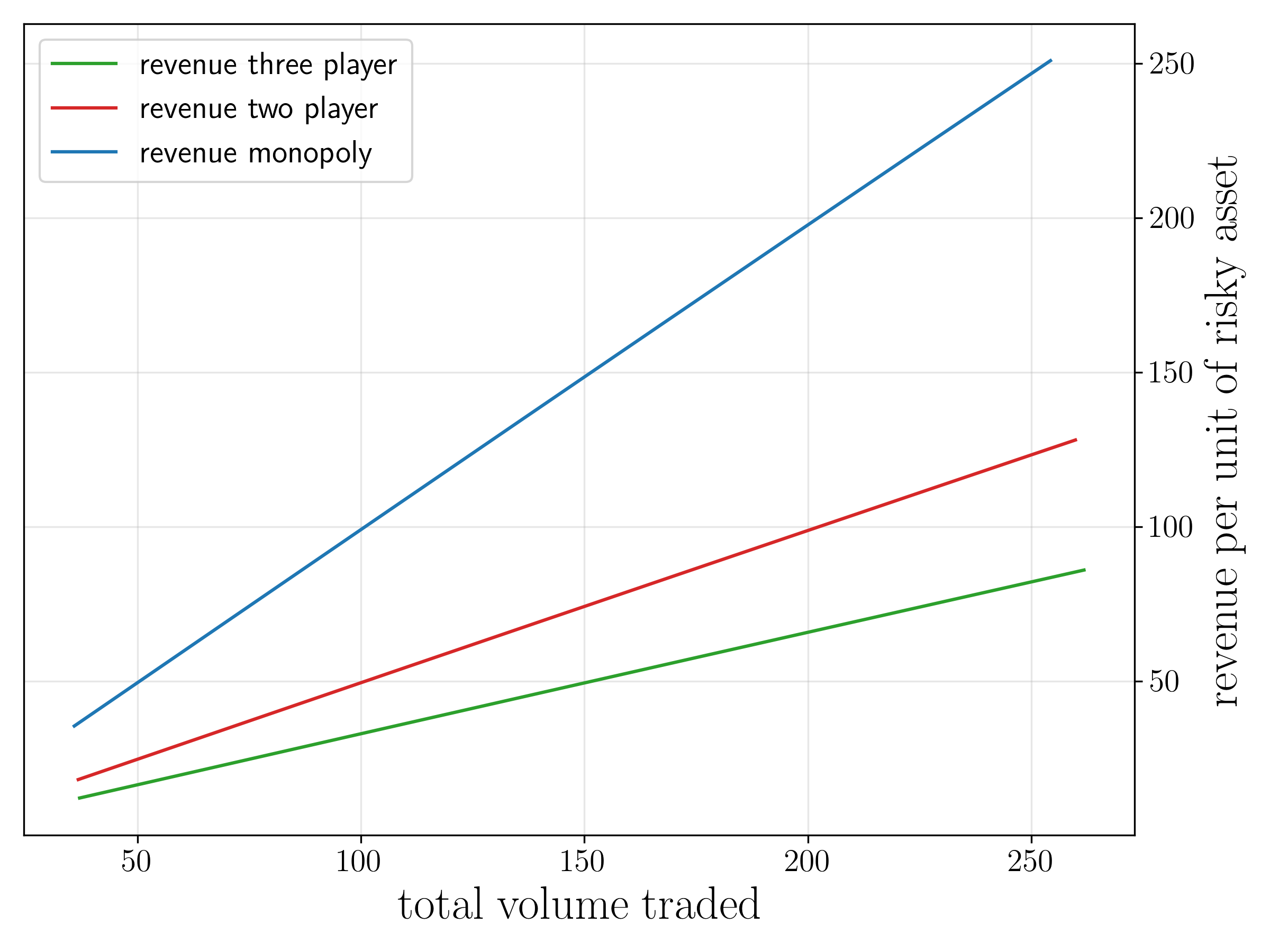}
        \caption{Revenue per player as a function of total traded volume in the monopoly (blue line), two-player (red line), and three-player (green line) scenarios, when liquidity is split among all participants.}
        \label{fig:rev-per-player-split-liq}
    \end{subfigure}
    \hspace*{0.01\textwidth}
    \begin{subfigure}[t]{0.48\textwidth}
        \centering
        \includegraphics[width=\textwidth]{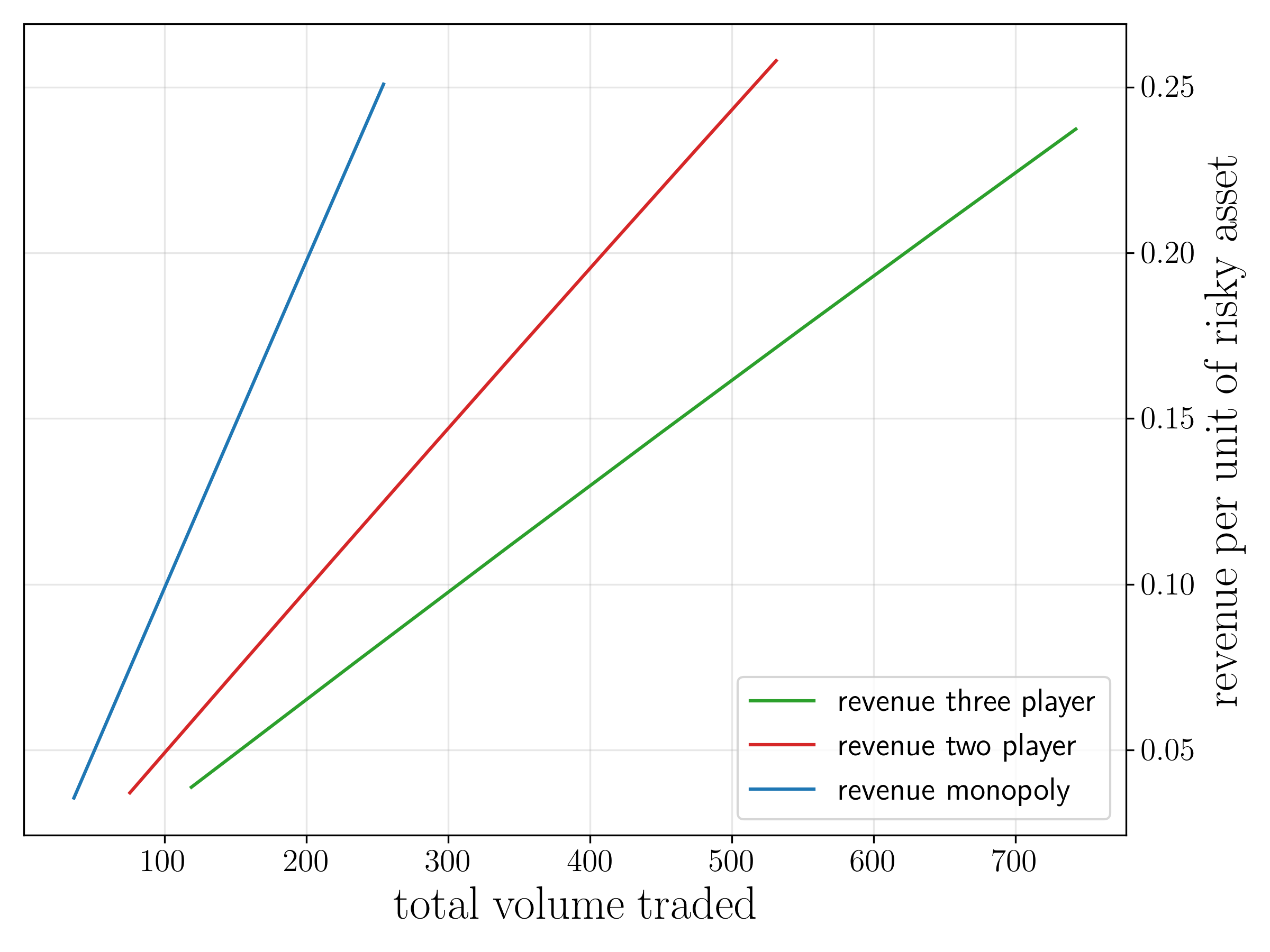}
        \caption{AMM total revenue as a function of total traded volume when all players provide the same amount of liquidity, in the monopoly (blue line), two-player (red line), and three-player (green line) scenarios.}
        \label{fig:amm-total-revenue-same-liq}
    \end{subfigure}
\end{figure}

\bibliography{thebib}

@article{baggiani2025optimal,
  title={Optimal Dynamic Fees in Automated Market Makers},
  author={Baggiani, Leonardo and Herdegen, Martin and S{\'a}nchez-Betancourt, Leandro},
  journal={arXiv preprint arXiv:2506.02869},
  year={2025}
}

@article{hasbrouck2022need,
  title={The need for fees at a dex: How increases in fees can increase dex trading volume},
  author={Hasbrouck, Joel and Rivera, Thomas J and Saleh, Fahad},
  journal={Available at SSRN 4192925},
  year={2022}
}

@inproceedings{milionis2024automated,
  title = {Automated Market Making and Arbitrage Profits in the Presence of Fees},
  author = {Milionis, Jason and Moallemi, Ciamac C and Roughgarden, Tim},
  booktitle = {International Conference on Financial Cryptography and Data Security},
  pages = {159--171},
  year = {2024},
  publisher = {Springer}
}

@article{milionis2022automated,
  title={Automated market making and loss-versus-rebalancing},
  author={Milionis, Jason and Moallemi, Ciamac C and Roughgarden, Tim and Zhang, Anthony Lee},
  journal={arXiv preprint arXiv:2208.06046},
  year={2022}
}

@inproceedings{milionis2022quantifying,
  title={Quantifying loss in automated market makers},
  author={Milionis, Jason and Moallemi, Ciamac C and Roughgarden, Tim and Zhang, Anthony Lee},
  booktitle={Proceedings of the 2022 ACM CCS Workshop on Decentralized Finance and Security},
  pages={71--74},
  year={2022}
}

@inproceedings{evans2021optimalfees,
  author = {Evans, Alex and Angeris, Guillermo and Chitra, Tarun},
  title = {Optimal Fees for Geometric Mean Market Makers},
  booktitle = {Financial Cryptography and Data Security. FC 2021 International Workshops},
  year = {2021},
  series = {Lecture Notes in Computer Science}
}

@article{he2024optimaldesignautomatedmarket,
      title={Optimal Design of Automated Market Makers on Decentralized Exchanges}, 
      author={Xue Dong He and Chen Yang and Yutian Zhou},
      journal={arXiv preprint arXiv:2404.13291},
      year={2024},
}

@article{bichuch2025price,
  title={The Price of Liquidity: Implied Volatility of Automated Market Maker Fees},
  author={Bichuch, Maxim and Feinstein, Zachary},
  journal={arXiv preprint arXiv:2509.23222},
  year={2025}
}

@article{bergault2026trading,
  title={Trading in CEXs and DEXs with Priority Fees and Stochastic Delays},
  author={Bergault, Philippe and Hafsi, Yadh and S{\'a}nchez-Betancourt, Leandro},
  journal={arXiv preprint arXiv:2602.10798},
  year={2026}
}

@article{cao2023structural,
  title={A structural model of automated market making},
  author={Cao, David and Kogan, Leonid and Tsoukalas, Gerry and Hemenway Falk, Brett},
  year={2023},
  journal={Available at SSRN 4591447},
  publisher={CBER CtCe Working Paper}
}

@inproceedings{fritsch2021note,
  title={A note on optimal fees for constant function market makers},
  author={Fritsch, Robin},
  booktitle={Proceedings of the 2021 ACM CCS Workshop on Decentralized Finance and Security},
  pages={9--14},
  year={2021}
}

@article{campbell2025optimal,
  title={Optimal fees for liquidity provision in automated market makers},
  author={Campbell, Steven and Bergault, Philippe and Milionis, Jason and Nutz, Marcel},
  journal={arXiv preprint arXiv:2508.08152},
  year={2025}
}

@article{nadkarni2024adaptive,
  title={Adaptive curves for optimally efficient market making},
  author={Nadkarni, Viraj and Kulkarni, Sanjeev and Viswanath, Pramod},
  journal={arXiv preprint arXiv:2406.13794},
  year={2024}
}

@article{avellaneda2008high,
  title={High-frequency trading in a limit order book},
  author={Avellaneda, Marco and Stoikov, Sasha},
  journal={Quantitative Finance},
  volume={8},
  number={3},
  pages={217--224},
  year={2008},
  publisher={Taylor \& Francis}
}

@book{cartea2015algorithmic,
  title={Algorithmic and high-frequency trading},
  author={Cartea, {\'A}lvaro and Jaimungal, Sebastian and Penalva, Jos{\'e}},
  year={2015},
  publisher={Cambridge University Press}
}

@book{gueant2016financial,
  title={The Financial Mathematics of Market Liquidity: From optimal execution to market making},
  author={Gu{\'e}ant, Olivier},
  year={2016},
  publisher={CRC Press}
}

@article{boyce2025market,
  title={Market making with exogenous competition},
  author={Boyce, Robert and Herdegen, Martin and S{\'a}nchez-Betancourt, Leandro},
  journal={SIAM Journal on Financial Mathematics},
  volume={16},
  number={2},
  pages={692--706},
  year={2025},
  publisher={SIAM}
}

@article{chilenje2025market,
  title={Market Making with Competition: A Stackelberg Equilibrium},
  author={Chilenje, Nehelo and Daba, Mandisa and Feleppa, Davide and Fellner, Cameron and S{\'a}nchez-Betancourt, Leandro},
  journal={Available at SSRN 5505480},
  year={2025}
}

@article{guo2025macroscopic,
  title={Macroscopic market making games},
  author={Guo, Ivan and Jin, Shijia},
  journal={Mathematical Finance},
  year={2025},
  publisher={Wiley Online Library}
}

@article{luo2021dynamic,
  title={Dynamic equilibrium of market making with price competition},
  author={Luo, Jialiang and Zheng, Harry},
  journal={Dynamic Games and Applications},
  volume={11},
  number={3},
  pages={556--579},
  year={2021},
  publisher={Springer}
}

@inproceedings{cartea2023execution,
  title={Execution and statistical arbitrage with signals in multiple automated market makers},
  author={Cartea, {\'A}lvaro and Drissi, Fay{\c{c}}al and Monga, Marcello},
  booktitle={2023 IEEE 43rd International Conference on Distributed Computing Systems Workshops (ICDCSW)},
  pages={37--42},
  year={2023},
  organization={IEEE}
}

@article{he2025arbitrage,
  title={Arbitrage on decentralized exchanges},
  author={He, Xue Dong and Yang, Chen and Zhou, Yutian},
  journal={arXiv preprint arXiv:2507.08302},
  year={2025}
}

@article{drissi2025equilibrium,
  title={Equilibrium Liquidity and Risk Offsetting in Decentralised Markets},
  author={Drissi, Fay{\c{c}}al and Wu, Xuchen and Jaimungal, Sebastian},
  journal={arXiv preprint arXiv:2512.19838},
  year={2025}
}

@article{jaimungal2023optimal,
  title={Optimal trading in automatic market makers with deep learning},
  author={Jaimungal, Sebastian and Saporito, Yuri F and Souza, Max O and Thamsten, Yuri},
  journal={arXiv preprint arXiv:2304.02180},
  year={2023}
}

@article{capponi2026optimal,
  title={Optimal Trading in Automated Market Makers},
  author={Capponi, Agostino and Coache, Anthony and Muhle-Karbe, Johannes},
  journal={Available at SSRN 6145286},
  year={2026}
}

@article{cartea2024strategic,
  title={Strategic bonding curves in automated market makers},
  author={Cartea, {\'A}lvaro and Drissi, Fay{\c{c}}al and S{\'a}nchez-Betancourt, Leandro and Siska, David and Szpruch, Lukasz},
  journal={Available at SSRN 5018420},
  year={2024}
}

@article{bayraktar2024dex,
  title={DEX specs: A mean field approach to DeFi currency exchanges},
  author={Bayraktar, Erhan and Cohen, Asaf and Nellis, April},
  journal={arXiv preprint arXiv:2404.09090},
  year={2024}
}

@article{baldacci2023mean,
  title={A mean-field game of market-making against strategic traders},
  author={Baldacci, Bastien and Bergault, Philippe and Possama{\"\i}, Dylan},
  journal={SIAM Journal on Financial Mathematics},
  volume={14},
  number={4},
  pages={1080--1112},
  year={2023},
  publisher={SIAM}
}
\bibliographystyle{plainnat}

\end{document}